\documentclass[12pt,preprint]{aastex}

\newcommand{\water}{${\rm H_2O}$} 
\newcommand{\methanol}{${\rm CH_3OH}$} 
\newcommand{\formalde}{${\rm H_2CO}$}
\newcommand{\sqcm}{${\rm cm^{-2}}$} 
\newcommand{\cubcm}{${\rm cm^{-3}}$} 
\newcommand{\mum}{${\rm \mu m}$} 
\newcommand{\kms}{${\rm km~s^{-1}}$} 
\newcommand{\thirteenco}{${\rm ^{13}CO}$}
\newcommand{\twelveco}{${\rm ^{12}CO}$} 
\newcommand{\eighteenco}{${\rm C^{18}O}$} 
\newcommand{\hcop}{${\rm HCO^+}$}
\newcommand{\bdop}{$b_{\rm D}$} 
\newcommand{\vlsr}{$V_{\rm LSR}$}

\usepackage{emulateapj5}

\slugcomment{To appear in ApJ 570 n2, 10 May 2002 (submitted 26 Sep
2001; accepted 7 Jan 2002)}

\shorttitle{The low mass protostar Elias 29}
\shortauthors{Boogert et al.}

\begin{document}

\title{The Environment and Nature of the Class I Protostar Elias~29:
Molecular Gas Observations and the Location of Ices}

\author{A.C.A. Boogert\altaffilmark{1}, 
        M.R. Hogerheijde\altaffilmark{2,3},
        C. Ceccarelli\altaffilmark{4},
        A.G.G.M. Tielens\altaffilmark{5,6},
        E.F. van Dishoeck\altaffilmark{7}, 
        G.A. Blake\altaffilmark{8}, 
        W.B. Latter\altaffilmark{9}, 
        F. Motte\altaffilmark{1}}

\altaffiltext{1}{California Institute of Technology, Downs Laboratory
              of Physics 320-47, Pasadena, CA 91125, USA;
              boogert@submm.caltech.edu}
\altaffiltext{2}{RAL, University of California at Berkeley, Astronomy
              Department, 601 Campbell Hall \# 3411, Berkeley, CA
              94720, USA}
\altaffiltext{3}{current address: Steward Observatory, University of
              Arizona, 933 N. Cherry Ave., Tucson, AZ 85721, USA}
\altaffiltext{4}{Observatoire de Bordeaux, B.P. 89, 33270 Floirac,
              France}
\altaffiltext{5}{Kapteyn Astronomical Institute, P.O. Box 800, 9700 AV
              Groningen, the Netherlands}
\altaffiltext{6}{SRON, P.O. Box 800, 9700 AV Groningen, the
              Netherlands}
\altaffiltext{7}{Leiden Observatory, P. O. Box 9513, 2300 RA Leiden,
              the Netherlands}
\altaffiltext{8}{California Institute of Technology, Division of
              Geological and Planetary Sciences 150-21, Pasadena, CA
              91125, USA}
\altaffiltext{9}{California Institute of Technology, SIRTF Science
              Center, IPAC, Pasadena, CA 91125, USA}

\begin{abstract}
  A (sub-)millimeter line and continuum study of the class I protostar
  Elias 29 in the $\rho$ Ophiuchi molecular cloud is presented, whose
  goals are to understand the nature of this source, and to locate the
  ices that are abundantly present along this line of sight. Within
  15--60$''$ beams, several different components contribute to the
  line emission. Two different foreground clouds are detected, an
  envelope/disk system and a dense ridge of \hcop--rich material.  The
  latter two components are spatially separated in millimeter
  interferometer maps.  We analyze the envelope/disk system by using
  inside-out collapse and flared disk models.  The disk is in a
  relatively face-on orientation ($\rm < 60^o$), which explains many
  of the remarkable observational features of Elias 29, such as its
  flat SED, its brightness in the near infrared, the extended
  components found in speckle interferometry observations, and its
  high velocity molecular outflow. It cannot account for the ices seen
  along the line of sight, however. A small fraction of the ices is
  present in a (remnant) envelope of mass 0.12--0.33 $M_{\odot}$, but
  most of the ices ($\sim$70\%) are present in cool ($T<$40 K)
  quiescent foreground clouds.  This explains the observed absence of
  thermally processed ices (crystallized H$_2$O) toward Elias 29.
  Nevertheless, the temperatures could be sufficiently high to account
  for the low abundance of apolar (CO, N$_2$, O$_2$) ices.  This work
  shows that it is crucial to obtain spectrally and spatially resolved
  information from single-dish and interferometric molecular gas
  observations in order to determine the nature of protostars and to
  interpret infrared ISO satellite observations of ices and silicates
  along a pencil beam.
\end{abstract}


\section{Introduction}~\label{se29:intro}

A rich chemical and physical interplay exists between gas and grains
in which molecules are formed on grains, creating ice mantles that are
preserved in environments ranging from quiescent dense molecular
clouds to envelopes and disks around protostars.  Various processes,
among which are bombardment by cosmic rays, ultraviolet irradiation,
heating, and shocks, can physically or chemically alter the icy
mantles, or return molecules into the gas phase (see \citealt{tiel97,
dish98}, and references therein).

A study of the chemical evolution of dense clouds to planet forming
disks would ideally involve observations of molecular gas and ices in
a range of environments, from quiescent clouds to disk dominated
protostars.  The most pristine, initial conditions are presumably well
sampled by field stars behind clouds tracing quiescent molecular cloud
material (e.g. \citealt{whit98}). Lines of sight to protostars are
more difficult to characterize, however, since they may trace
quiescent foreground material, in addition to the gas and ices in
their envelopes and disks (e.g. \citealt{boog00b}). It is thus crucial
to characterize the line of sight conditions in order to locate the
ices and derive physical conditions and eventually the evolution of
the molecular gas and solid state in the interstellar medium.

\begin{table*}[t!]
\center
{\footnotesize
\caption{Observational Summary}~\label{t:obssum}
\begin{tabular}{lllll}
\tableline
\noalign{\smallskip} 
Telescope    & Beam $\varnothing$ (\arcsec) & $\lambda$ or $\nu$      & Species$^a$	   & Date or Reference \\
\noalign{\smallskip}  
\tableline
\noalign{\smallskip} 
NRAO-12m     & 43--86	      & 70--145 GHz	    & CO, CS, \hcop        & 05/1995 \\
	     &		      &	    		    & \formalde, \methanol &	   \\
JCMT	     & 14--22	      & 200--400 GHz	    & CO, CS, \hcop        & 1995-1997 \\
	     &		      &	    		    & \formalde, \methanol &	   \\
JCMT	     & 7	      &	692 GHz	    	    & CO 6-5 		   & Ceccarelli et al., in prep. \\
CSO	     & 21--35	      &	200--400 GHz	    & \formalde, \methanol & 1999-2001 \\
	     &		      &	    		    & CO, \hcop\ maps      &	       \\ 
OVRO	     & 4$\times$8, 3$\times$6 & 87, 110 GHz         & CO, HCO$^+$, SiO     & 09/1999-07/2000 \\
OVRO	     & 4$\times$8, 3$\times$6 & 2.7, 3.3 mm         & Continuum	      	   & 09/1999-07/2000 \\
IRAM-30m     & 15	      & 1.3 mm		    & Continuum		   & \citealt{mott98} \\
ISO SWS      & 14--33	      & 2.3--45 \mum\	    & Continuum     	   & \citealt{boog00b} \\
             &                &                	    & Ices, Silicates	   &                   \\
ISO LWS	     & 80             & 45--200 \mum\	    & Continuum	   	   & \citealt{boog00b} \\
\noalign{\smallskip} 
\tableline
\noalign{\smallskip}
\multicolumn{5}{l}{$\rm ^a$ main species and/or isotopes measured. For more details see Table~\ref{t:sdish}.}\\
\end{tabular}
}
\end{table*}

In this paper, we study the line of sight of the class I protostar
Elias 29 in the $\rho$ Oph molecular cloud, using (sub)millimeter
single dish and interferometer gas phase observations.  This object is
one of the most luminous protostars ($\sim 36~L_{\odot}$;
\citealt{chen95}) in the nearby $\rho$ Oph complex ($d\sim 160$ pc;
\citealt{whit74}), yet little is known about its nature and line of
sight conditions.  Abundant ice has been detected in its direction
\citep{zinn85, tana90}. A detailed analysis of ice band profiles
indicates that the ices are not strongly thermally processed (i.e. the
ices are not crystallized or segregated), despite the presence of
abundant warm molecular gas toward the object \citep{boog00b}.  This
contrasts with high mass, luminous ($\rm >10^4~L_{\odot}$) protostars,
where significant thermal processing of the ices accompanies the
presence of abundant warm molecular gas \citep{boog00a, tak00}.  The
result obtained for Elias 29 can only be understood once the location
of the ices and the physical conditions of the various gas components
are known.  Therefore, in this paper we try to identify any foreground
material, the presence of a circumstellar envelope as well as the
presence and orientation of a circumstellar disk, and the column
density of each component.  We will then address the question where
the ices are located, and what their relation is to the young star.
This study will, as a consequence, reveal important information on the
nature and evolutionary stage of Elias 29, which has many interesting
and unique properties \citep{elia78}.

Details of the single dish and interferometer observations are
presented in \S 2, and the maps and spectra are decomposed and
interpreted in \S 3.  The physical conditions are determined for the
different components along the line of sight, among which are two
foreground clouds (\S 4.1), a remnant envelope and face-on disk (\S
4.2), as well as a dense ridge from which Elias 29 probably formed (\S
4.3). The gas phase conditions and abundances are linked to the ice
observations in \S 4.4. The depletion of gas phase species is compared
to young class 0 objects and quiescent clouds in \S 4.5. The results
are summarized in \S 5.

\section{Observations}

Single dish and interferometer millimeter wave spectral line and
continuum observations were made toward the infrared position of Elias
29 ($\alpha$ = 16$\rm ^h$27$\rm ^m$09$\rm ^s$.5, $\delta$ = $-24\rm
^o$37$'$18$''$; J2000). Rotational lines of CO, \hcop, CS, \formalde,
\methanol, and isotopes were selected in the 70-400 GHz spectral range
based on their sensitivity to column density, temperature, and density
\citep{blak95}. Below we discuss the technical details of these
observations.  Not discussed is a map of the CO 6-5 line (692 GHz),
for which we refer to C. Ceccarelli et al. (in preparation).  Also,
the technical details of the 1.3 mm IRAM-30m continuum map at
15\arcsec\ spatial resolution that is used in this paper are discussed
elsewhere (Motte, Andr\'e, \& Neri 1998).  Finally, the infrared
spectral energy distribution (SED) that we use consists of 2-45, and
45-200 \mum\ spectra that were obtained with the ISO--SWS and LWS
instruments in apertures of $\sim$25\arcsec\ and $\sim$80\arcsec\
respectively, and were extensively discussed in \citet{boog00b}. A
summary of all the observations used in this paper is given in
Table~\ref{t:obssum}.

\subsection{Singe Dish 70-400 GHz Spectral Line Observations}

Single dish NRAO-12m, JCMT, and CSO millimeter wave observations were
made with a single pointing toward Elias 29 during a number of
observing runs in the 1995-2001 period (Table~\ref{t:obssum}). In some
lines, small maps were made of the Elias 29 environment with the CSO.
At the JCMT and CSO, we used the DAS and AOS spectrometer back-ends in
the highest available spectral resolution mode ($\sim$0.1 \kms), in
the 200-400 GHz spectral range. At the NRAO-12m, we used the
autocorrelator back-end, with a channel width of 0.049 MHz, resulting
in Nyquist sampled spectra with resolutions of 0.40-0.20 \kms\ at the
frequencies 70-145 GHz.

For low frequency transitions, that trace extended cloud structure, we
took $\alpha$ = 16$\rm ^h$23$\rm ^m$01$\rm ^s$.5, $\delta$ = $-24\rm
^o$36$'$58$''$ (J2000) as an off position, which is found to have very
little or no \thirteenco\ emission \citep{lore89}. For higher
frequency CO transitions, we took an off position of 2700\arcsec\ in
azimuth, and for other molecules the azimuth offset was 900\arcsec. No
contamination by emission in the off position is evident in our
spectra.

\begin{table*}
\center
{\footnotesize
\caption{Applied Beam Efficiencies}~\label{t:eta}
\begin{tabular}{lll}
\tableline
\noalign{\smallskip} 
Telescope &          Frequency       & $\eta_{\rm mb}$$\rm ^a$\\
          &          GHz             &                \\
\noalign{\smallskip}  
\tableline
\noalign{\smallskip} 
NRAO-12m         & 70-85  & 1.00        \\
         	 & 85-90  & 0.95        \\
         	 & 95-115 & 0.93        \\
         	 & 140-150& 0.81        \\
JCMT         	 & 200-270& 0.68        \\
         	 & 320-380& 0.60        \\
CSO          	 & 200-270& 0.76$\rm ^b$\\
         	 & 320-380& 0.78$\rm ^b$\\
\noalign{\smallskip} 
\tableline
\noalign{\smallskip}
\multicolumn{3}{l}{$\rm ^a~\eta_{\rm m}^*$ given for NRAO-12m}\\
\multicolumn{3}{l}{$\rm ^b$ for 03/1999; 10 \% lower in 01/2001}\\
\end{tabular}
}
\end{table*}

Our line observations were corrected for atmospheric attenuation and
telescope losses, using the standard chopping wheel technique
\citep{kutn81}. The NRAO-12m data retrieved from the telescope are on
a $T_{\rm R}^*$ scale, and the CSO and JCMT data in $T_{\rm A}^*$. To
convert to the main beam brightness temperature ($T_{\rm mb}$), we
divide by the beam efficiencies:

\begin{eqnarray}
T_{\rm mb} & = & {T_{\rm R}^*}/{\eta_{\rm m}^*} \ \ \ \ {\rm (NRAO-12m)}\\
           &   & {T_{\rm A}^*}/{\eta_{\rm mb}}  \ \ \ {\rm (JCMT,CSO).}
\end{eqnarray}

 This corrects for losses in the beam side lobes. For the JCMT and
CSO, losses due to forward spillover and scattering are included in
$\eta_{\rm mb}$.  The beam efficiencies applied to our data are
summarized in Table~\ref{t:eta}.  For the CSO telescope\footnote{CSO
beam parameters are available at
\url{http://www.submm.caltech.edu/cso/receivers/beams.html}},
$\eta_{\rm mb}$ and the main beam size (Table~\ref{t:sdish}) were
determined by observing Mars during observing runs in March 1999 and
January 2001. For the NRAO-12m\footnote{User's Manual for the NRAO-12m
Millimeter-Wave Telescope \citep{mang99} is available at
\url{http://www.tuc.nrao.edu/12meter/obsinfo.html}}, we used the
theoretical values of $\eta_{\rm m}^*$ from the Ruze equation,
increased by a small factor ($\sim 1.08$), needed to reproduce the
experimental values \citep{mang99}. As a check, we determined
$\eta_{\rm m}^*$ from our observation of Jupiter, yielding $\eta_{\rm
m}^*$=0.80 at 145 GHz, in excellent agreement with the assumed value.
For the JCMT, the efficiency factors are taken from regularly measured
values reported on the internet\footnote{JCMT beam parameters are
available at \url{http://www.jach.hawaii.edu/JACpublic/JCMT/}}.

The calibration accuracy between different observing runs is not
expected to be better than 25\% (e.g. \citealt{mang93}).  As a check
on the calibration accuracy, we observed the very bright nearby source
IRAS 16293--2422 during our JCMT and CSO runs.  We find that line
intensities, observed with the same telescope during the night and in
separate observing runs, indeed scatter with 20-30\% variations around
the mean. Similar differences are found when comparing JCMT
observations to the standard spectra available at the JCMT internet
page.  In this paper, we will assume an uncertainty of 25\% in the
line intensity, unless noted otherwise.

The data were reduced with the GILDAS/CLASS reduction package,
applying low order baselines and using weighted means in averaging
individual spectra. The reduced single dish spectra and maps are
presented in Figs.~\ref{f:hcop}--~\ref{f:sdish} and
Table~\ref{t:sdish}, and analyzed in \S 3.1.

\subsection{Interferometer Spectral Line and Continuum Observations}

Elias 29 was observed with the OVRO millimeter array in the 1999/2000
season (Table~\ref{t:obssum}). The low elevation of the source allowed
for only short tracks ($\sim$7 hours).  The digital correlator system
was used in the ``1~8~2'' mode, providing simultaneous observations of
the \eighteenco\ 1-0 and \thirteenco\ 1-0 lines in one local
oscillator setting as well as the \hcop\ 1-0 and SiO 2-1 lines in
another.  The spectral resolution is 0.35 \kms, with a velocity
coverage of 20 \kms.  The telescope configurations L, H, and E were
used in all lines, as well as the continuum. The weather conditions
were best during the L and H observations, resulting in $T_{\rm
sys}\sim$ 500 K at 110 GHz, and 800 K at 87 GHz. System temperatures
were worse in the E track: 700 and 1300 K respectively.  The data were
reduced in a standard way in the OVRO/MMA reduction package.  The
nearby source NRAO 530 was used as a gain calibrator, and flux
calibration was performed on the planets Uranus and Neptune. The
calibrated tracks were combined, and deconvolved images were
constructed with the MIRIAD software package using uniform weighting.
The resulting synthesized elliptical beam sizes are 4$\times$8\arcsec\
at 87 GHz and 3$\times$6\arcsec\ at 110 GHz; the noise level achieved
per channel in the final data is 0.26 Jy/beam and 0.13 Jy/beam
respectively.  Strong and extended \thirteenco\ and \hcop\ 1-0
emission is detected, with perhaps a weak detection of \eighteenco\
1-0 (Figs.~\ref{f:ovromap}--\ref{f:ovrochan}). SiO 2-1 is undetected
toward Elias~29.

No obvious strong continuum source was seen in the OVRO 2.7 and 3.4 mm
maps. A weak 3 $\sigma$ peak (7 mJy) is detected at 3.4 mm. This is
the strongest peak in the map, and it is centered on the infrared
position of Elias 29.  At 2.7 mm a peak of only 2 $\sigma$ (5 mJy)
significance appears at the same position.  For our analysis, we take
a weighted mean of these values, which we will refer to as the
continuum flux at 3.0 mm: $F$(3 mm)=6.1$\pm$1.7 mJy. This is only a
3.5 $\sigma$ result, and needs confirmation by additional
observations.

\section{Results}

\subsection{Large Scale Emission: Single Dish Spectra}

\begin{figure*}[t!]
\center
\includegraphics[angle=90, scale=0.60]{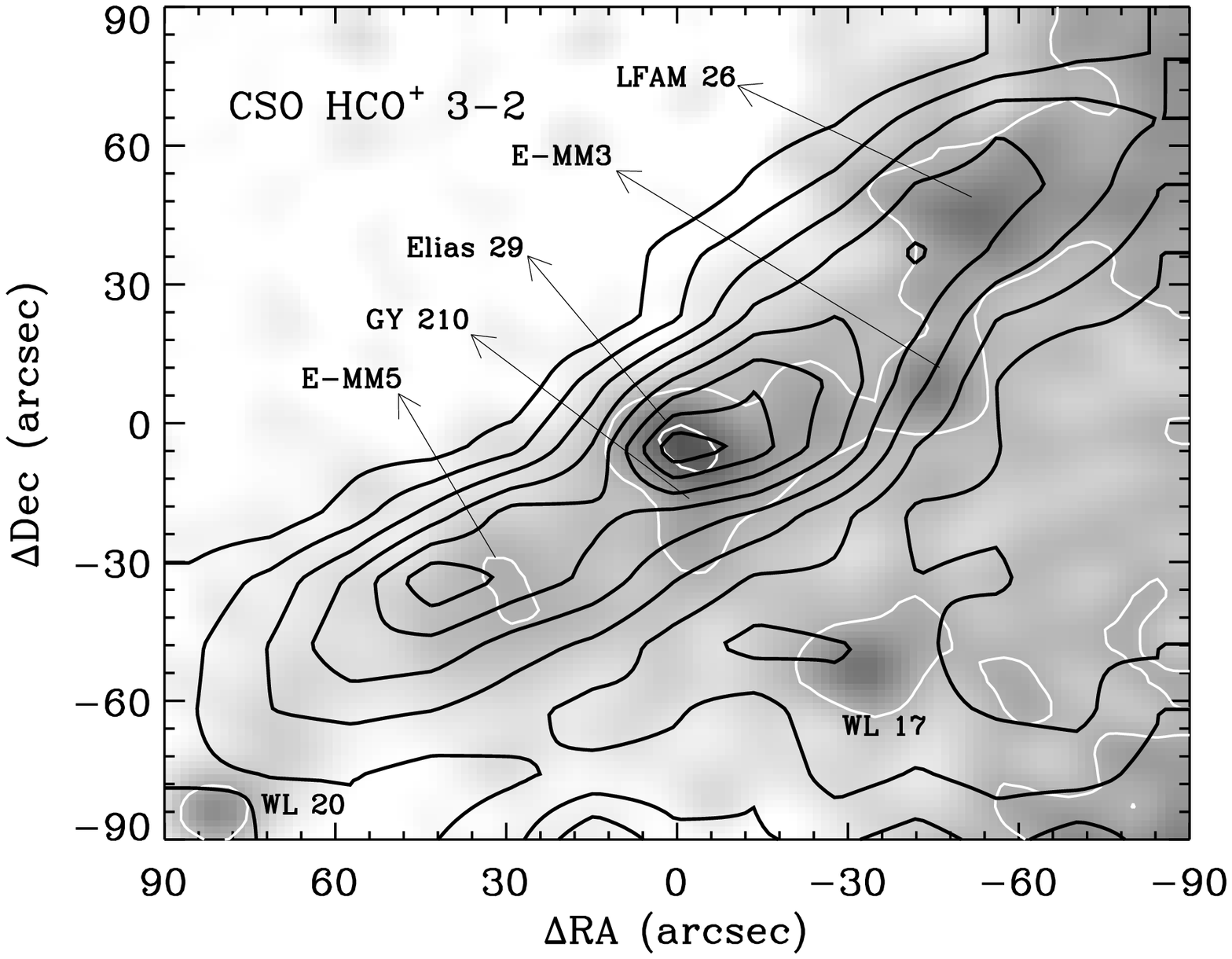} 
\caption{Overlay of IRAM-30m 1.3 mm continuum map (gray scale and two
white contours; \citealt{mott98}) and integrated CSO \hcop\ 3-2
emission (black contours) of the Elias 29 region, with source
identifications. Contour levels for the continuum map are at the 7,
and 14 $\sigma$ confidence level, with $\sigma$ = 10 mJy/15$''$ beam,
and $\int T_{\rm MB}~dv$ = 1.1, 1.6, ..., 4.6 K.\kms\ for \hcop\ 3-2
($\sigma$ = 0.2 K.\kms). The 1.3 mm map was aligned to the radio
continuum positions of Elias 29 and WL~20. The \hcop\ map in turn was
aligned to the 1.3 mm map, for which a small shift of 5\arcsec\ to the
southwest was needed.}~\label{f:hcop}
\end{figure*}

\begin{figure*}[b]
\center
\includegraphics[angle=0, scale=0.50]{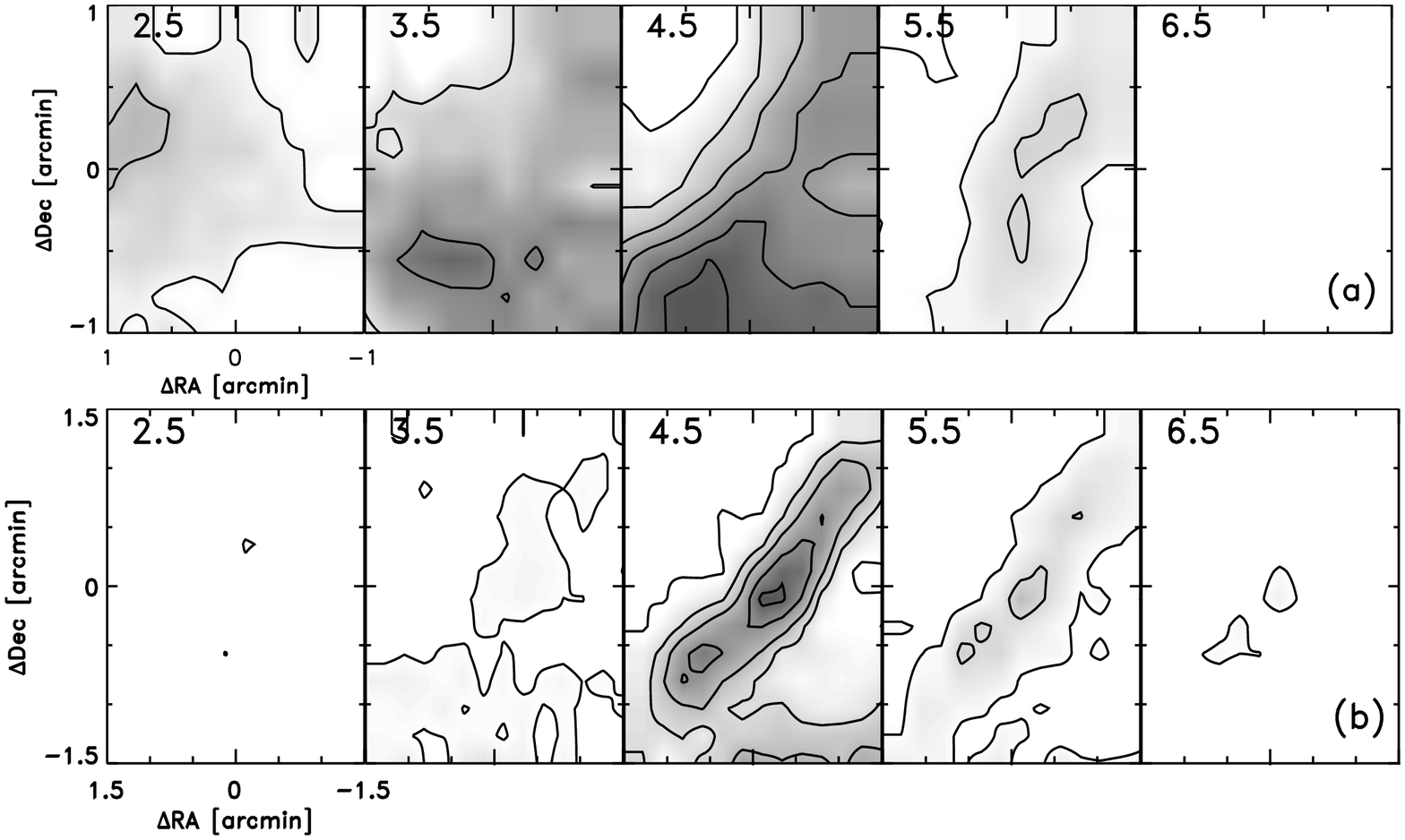} 
\caption{Channel maps of \eighteenco\ 2-1 (a) and \hcop\ 3-2
(b) emission in the Elias 29 region, observed with the CSO
telescope.  The emission is integrated over a 1~\kms\ interval, with
the central velocity (in \kms) indicated in the upper left
corners. The contour levels are equally spaced $\int T_{\rm MB}~dv$ =
0.8,2.1,...9.9 K.\kms\ for \eighteenco\ ($\sigma =0.20$ K.\kms), and
0.3, 0.9, ..., 2.7 K.\kms\ for \hcop\ ($\sigma =0.09$ K.\kms). Note
that the map scale is 2$\times$2 arcmin for \eighteenco, and
3$\times$3 arcmin for \hcop. Foreground clouds are seen in
\eighteenco\ 2-1 at 2.5 and 3.7 \kms. The Elias 29 core and
surrounding ridge are seen at higher velocities, especially in
\hcop.}~\label{f:csochan}
\end{figure*}

The single dish maps show that Elias 29 is not a particularly prominent
center of molecular line emission.  Within a radius of $\sim 1.5'$,
the \hcop\ 3-2 and \eighteenco\ 2-1 emission is highly structured, and
quite differently distributed (Figs. \ref{f:hcop} and
\ref{f:csochan}).

The \hcop\ 3-2 emission, tracing high densities ($\geq 10^5$ \cubcm),
is concentrated in a remarkable ridge-like structure, oriented in the
southeast-northwest direction (Fig.~\ref{f:hcop}), at a velocity of
\vlsr$\sim$5.0 \kms\ (Fig.~\ref{f:csochan}).  It is likely no
coincidence that the protostars Elias 29, WL 20, LFAM 26, GY~210 as
well as the 1.3 mm continuum protostellar condensations E-MM3 and
E-MM5 are all located along this dense ridge (Fig.~\ref{f:hcop}). The
ridge is also particularly prominent in the 800 \mum\ continuum
\citep{john00b}. Star formation along dense filamentary structures is
common in the $\rho$ Oph cloud, and has been explained by the presence
of magnetic field tubes or, more likely, by externally induced shocks
(see \citealt{mott98} for a short discussion).

\begin{figure*}[p]
\center
\includegraphics[angle=270, scale=0.80]{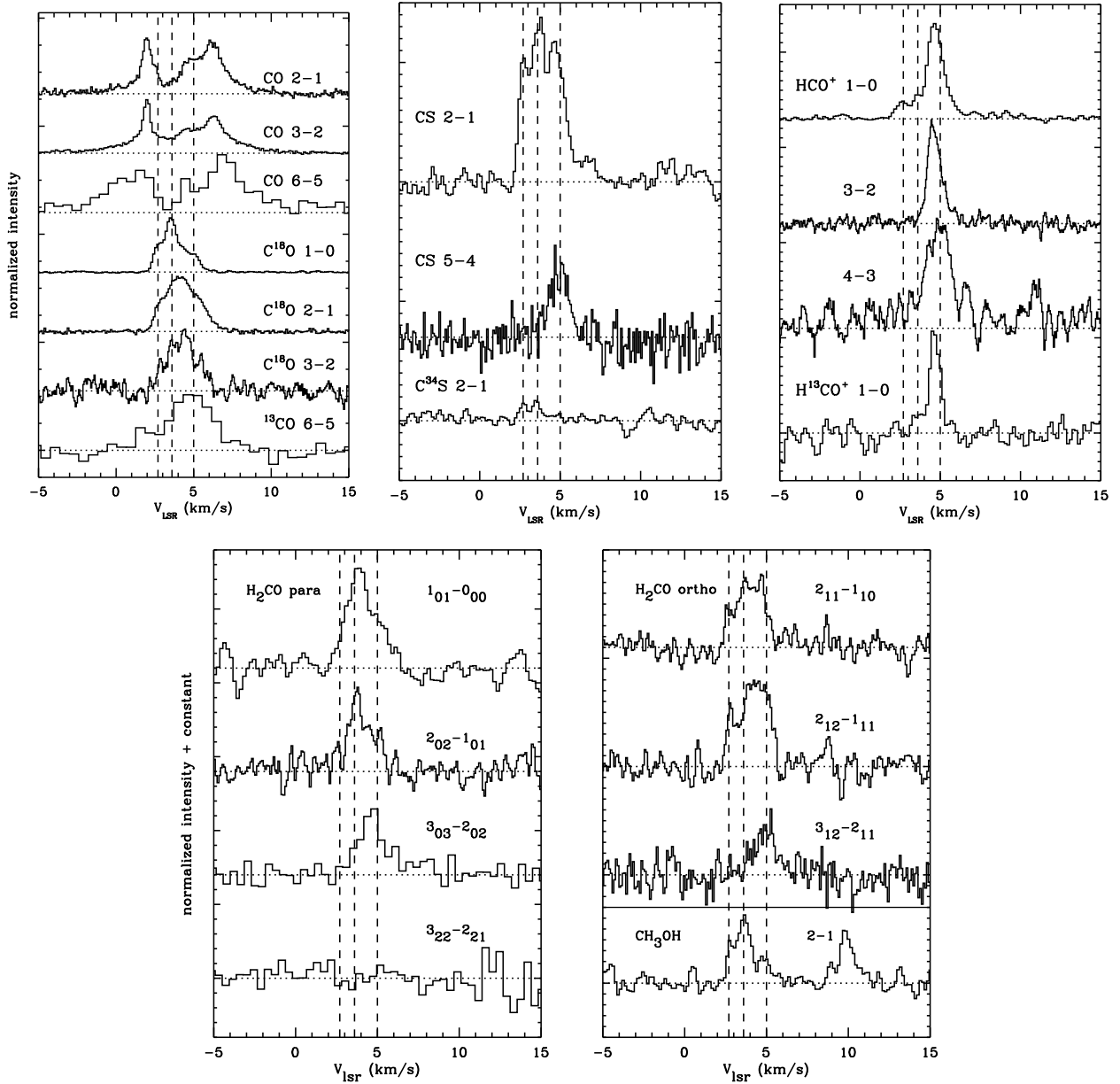} 
\caption{Emission lines of CO, CS, \hcop, \formalde, and \methanol\
toward Elias 29.  Vertical dashed lines indicate different dynamic
components at \vlsr=2.7, 3.8 and 5.0 \kms. The spectra have been
arbitrarily scaled and shifted along the intensity axis.  See
Table~\ref{t:sdish} for line intensities.}~\label{f:sdish}
\end{figure*}

The \eighteenco\ 2-1 emission, a column rather than volume density
tracer, shows that at least three clouds are present along the line of
sight of Elias 29 (Fig.~\ref{f:csochan}). The channel maps show a
cloud at \vlsr$\sim 2.7$ \kms\ that peaks to the northeast of
Elias 29, and a cloud at \vlsr$\sim 3.8$ \kms\ spread rather
evenly over the map. The brightest cloud at $\sim$5 \kms\ peaks
prominently near the south-southwest of the map, and is probably that
in which the dense \hcop\ ridge resides, given the similar velocities.
All these clouds are likely present in the foreground, since
absorption in the \twelveco\ emission lines is seen at these
velocities (Fig.~\ref{f:sdish}).

\begin{table*}[p]
{\footnotesize
\caption{Rotational Transitions Observed toward Elias~29 (Center
Position) with Single Dish Telescopes\label{t:sdish}}
\begin{tabular}{llrcccccll}
\tableline
\noalign{\smallskip} 
{Molecule} & {Transition} & {Frequency} & {$T_{\rm MB}$\tablenotemark{a}} &
  {$\int T_{\rm MB}~dv$\tablenotemark{a}} & {FWHM} & {\vlsr} & {Beam $\varnothing$} &
  {Telescope} & {Date} \\
{} & {} & {MHz} & {K} & {K.\kms} & {\kms} & {\kms} & {arcsec} & 
  {}& {}\\
\tableline
\noalign{\smallskip} 
CO              & 2-1           & 230538.0  & 15.9         & 19.4                 & 6    & 2/6.5         & 21   & JCMT  & 03/1997 \\
                & 3-2           & 345796.0  & 26           & 98                   & 7    & 2/6.5         & 14   & JCMT  & 02/1996 \\
                &               &           & 9            & 46                   & 6    & 2/6.5         & 21   & CSO   & 07/2000 \\
                & 6-5\tablenotemark{b}& 691473.0  & 17           & 124                  & 12   & 1.8/6.5       & 7    & JCMT  & 04/1995 \\
\noalign{\smallskip} 
$^{13}$CO       & 6-5\tablenotemark{b}& 661067.4  & 10           & 39                   & 3.6  & 4.78          & 7    & JCMT  & 04/1995 \\
\noalign{\smallskip} 
C$^{17}$O       & 1-0 triplet   & 112358.7  & 0.40         & 0.99                 & 2.3  & \nodata       & 56   & NRAO  & 05/1995 \\
                &               & 112359.0  & 0.94         & 1.00                 & 1.0  & 3.58          &      & NRAO  & 05/1995 \\
                &               & 112360.0  & 0.68         & 1.05                 & 1.5  & \nodata       &      & NRAO  & 05/1995 \\
                & 2-1 multiplet & 224714.3  & 1.76         & 4.95                 & 2.63 & 4.08          & 22   & JCMT  & 03/1995 \\
\noalign{\smallskip} 
C$^{18}$O       & 1-0           & 109782.2  & 6.34         & 11.0                 & 2.0  & 3.6           & 57   & NRAO  & 05/1995 \\
                & 2-1           & 219560.4  & 5.81         & 16.2                 & 2.62 & 4.19          & 22   & JCMT  & 03/1995 \\
                &               &           & 9.6          & 21.3                 & 2.41 & 4.05          & 35   & CSO   & 01/2001 \\
                & 3-2           & 329330.6  & 4.3          & 10.4                 & 2.28 & 4.24          & 15   & JCMT  & 02/1996 \\
\noalign{\smallskip} 
CS              & 2-1           & 97981.0   & 1.28         & 3.55                 & 2.61 & 3.96          & 64   & NRAO  & 05/1995 \\
                & 5-4           & 244935.6  & 0.56         & 0.94                 & 1.56 & 4.92          & 20   & JCMT  & 03/1995 \\
                & 7-6           & 342883.0  & $<$0.09      & $<0.13$              & \nodata & \nodata    & 14   & JCMT  & 02/1996 \\
\noalign{\smallskip} 
C$^{34}$S       & 2-1           & 96412.9   & 0.14         & 0.23                 & 1.62 & 3.29          & 65   & NRAO  & 05/1995 \\
\noalign{\smallskip} 
H$_2$CO para    &$1_{01}-0_{00}$& 72838.0   & 0.62         & 1.44                 & 2.19 & 3.80/5.39     & 86   & NRAO  & 05/1995 \\
                &$2_{02}-1_{01}$& 145603.0  & 0.56         & 0.89                 & 2.0  & 3.67/4.10     & 43   & NRAO  & 05/1995 \\
                &$3_{03}-2_{02}$& 218222.2  & 0.39         & 0.66                 & 1.89 & 4.60          & 22   & JCMT  & 03/1995 \\
                &$3_{22}-2_{21}$& 218475.6  & $<$0.06      & \nodata              & \nodata  & \nodata   & 22   & JCMT  & 03/1995 \\
\noalign{\smallskip} 
H$_2$CO ortho   &$2_{12}-1_{11}$& 140839.5  & 0.80         & 1.85                 & 2.17 & 4.20          & 45   & NRAO  & 05/1995 \\
                &$2_{11}-1_{10}$& 150498.4  & 0.80         & 1.85                 & 2.17 & 4.20          & 42   & NRAO  & 05/1995 \\
                &$3_{12}-2_{11}$& 225697.8  & 0.41         & 0.76                 & 1.76 & 4.93          & 21   & JCMT  & 03/1995 \\
                &               &           & 0.26         & 0.48                 & 1.70 & 4.51          & 32   & CSO   & 07/2000 \\
\noalign{\smallskip} 
CH$_3$OH        &2-1 triplet    & 96739.4   & 0.15         & 0.19                 & 1.24 & \nodata       & 65   & NRAO  & 05/1995 \\
                &               & 96741.4   & 0.18         & 0.33                 & 1.68 & 3.59          &      & NRAO  & 05/1995 \\

                &               & 96744.6   & $<$0.03      & \nodata              &\nodata&\nodata       &      & NRAO  & 05/1995 \\
                &5-4 multiplet  & 241802.0  & $<$0.05      & \nodata              &\nodata&\nodata       & 20   & JCMT  & 03/1997 \\
                &               &           & $<$0.02      & \nodata              &\nodata&\nodata       & 31   & CSO   & 06/2000 \\
\noalign{\smallskip} 
\hcop\          &1-0            & 89188.5   & 3.57         & 5.69                 & 3.3/1.0& 4.3/4.7     & 70   & NRAO  & 05/1995 \\
                &3-2            & 267557.6  & 1.06         & 2.7                  & 2.28 & 4.77          & 18   & JCMT  & 05/1996 \\
                &               &           & 4.34         & 4.9                  & 1.05 & 4.63          & 28   & CSO   & 06/2000 \\
                &4-3            & 356734.3  & 0.30         & 0.80                 & 2.5  & 4.5           & 21   & CSO   & 03/1999 \\
\noalign{\smallskip} 
H$^{13}$CO$^+$  &1-0            & 86754.3   & 0.38         & 0.36                 & 0.90 & 4.66          & 72   & NRAO  & 05/1995 \\
                &3-2            & 260255.5  & 0.04 (0.02)  & 0.09 (0.02)          & 2.15 & 4.89          & 29   & CSO   & 03/1999 \\
\noalign{\smallskip} 
\tableline
\noalign{\smallskip} 
\tablenotetext{a}{Calibration errors are 25\%, unless noted otherwise in parentheses.}
\tablenotetext{b}{CO 6-5 spectra presented and analyzed in detail in C. Ceccarelli et al., in prep.}
\end{tabular}
}
\end{table*}

\begin{table*}[t!]
\center
{\footnotesize
\caption{Integrated Intensities of Decomposed Lines at Center Position}~\label{t:decomp}
\center
\begin{tabular}{lllll}
\tableline
\noalign{\smallskip} 
Molecule  & Transition    & \multicolumn{3}{c}{$\int T_{\rm MB}~dv$ for velocity component$\rm ^a$}\\
          &               & 2.7~\kms    & 3.8~\kms      & 5.0~\kms                            \\
\noalign{\smallskip}  
\tableline
\noalign{\smallskip} 
C$^{18}$O       & 1-0           & 1.78          & 5.18          & 3.97          \\
                & 2-1           & 1.93          & 4.19          & 9.50          \\
                & 3-2           & 1.00          & 2.75          & 6.78          \\
\noalign{\smallskip} 
C$^{17}$O$^{\rm b}$ & 1-0               & 0.66          & 1.94          & $<$ 1.6       \\
\noalign{\smallskip} 
CS              & 2-1           & 0.69          & 1.01          & 1.72          \\
                & 5-4           & $<$ 0.07      & $<$ 0.09      & 0.82          \\
                & 7-6           & $< 0.13$      & $< 0.13$      & $<$ 0.13      \\
\noalign{\smallskip} 
C$^{34}$S       & 2-1           & 0.09          & 0.12          & $<$ 0.05      \\
\noalign{\smallskip} 
H$_2$CO para    &$1_{01}-0_{00}$& $< 0.15$      & 0.55          & 0.65          \\
                &$2_{02}-1_{01}$& $<$ 0.07      & 0.43          & 0.39          \\
                &$3_{03}-2_{02}$& $< 0.03$      & $< 0.04$      & 0.82          \\
                &$3_{22}-2_{21}$& $< 0.03$      & $< 0.04$      & $< 0.12$      \\
\noalign{\smallskip} 
H$_2$CO ortho   &$2_{12}-1_{11}$& 0.30          & 0.50          & 1.03          \\
                &$2_{11}-1_{10}$& 0.21          & 0.52          & 0.65          \\
                &$3_{12}-2_{11}$& $< 0.06$      & $<$ 0.09      & 0.78          \\
\noalign{\smallskip} 
CH$_3$OH        &2-1 96739.4    & 0.03          & 0.14          & $<$ 0.05      \\ 
                &2-1 96741.4    & 0.06          & 0.18          & 0.10          \\
                &2-1 96744.6    & $< 0.02$      & $< 0.02$      & $< 0.05$      \\
\noalign{\smallskip} 
\hcop\          &1-0            & 0.49          & 0.46          & 4.26          \\
                &3-2            & $< 0.04$      & $< 0.07$      & 3.17          \\
                &4-3            & $< 0.05$      & $< 0.08$      & 0.76          \\
\noalign{\smallskip} 
H$^{13}$CO$^+$  &1-0            & $< 0.02$      & 0.05          & 0.33          \\
\noalign{\smallskip} 
\tableline
\multicolumn{5}{p{10cm}}{$^{\rm a}$ Relative uncertainties are smaller
than the error in the absolute calibration (25\%)}\\
\multicolumn{5}{p{10cm}}{$^{\rm b}$ Derived from 112360.0 MHz fine structure line, 
multiplied by a factor of 3 to correct for emission in other fine structure lines.}\\
\noalign{\smallskip}
\end{tabular}}
\end{table*}

In various other single dish lines these clouds show up as discrete
emission components (Fig.~\ref{f:sdish}).  The 2.7 and 3.8 \kms\
components are primarily seen in the low-lying transitions of the CS,
\hcop, \formalde, and \methanol\ molecules, indicating low densities
and temperatures of these extended clouds. The emission at 5.0 \kms,
identified with the Elias 29 core and ridge, shows up in the high
excitation lines.

Given the fact that these emission components have different spatial
distributions, their relative intensity also depends on beam size.
The \hcop\ 3-2 line is a factor of 4 stronger and a factor of 2
narrower in the CSO data compared to JCMT: the larger CSO beam (28$''$
versus 18$''$; Table~\ref{t:sdish}) picks up more emission from the
ridge, which is bright in the west (\S 3.2).  The wide beam NRAO-12m
\hcop\ 1-0 and H$^{13}$CO$^+$ 1-0 spectra have remarkably narrow
5.0~\kms\ components, originating from the extended ridge.  In
contrast, the \formalde\ $3_{12}-2_{11}$ line may peak on the Elias 29
core, rather than the ridge.  It has the same width in the CSO and
JCMT beams, but the CSO spectrum is weaker (Table~\ref{t:sdish}).

We have performed a Gaussian decomposition of the emission lines to
separate the foreground clouds from the dense material around Elias
29.  The derived relative strength of these blended lines depends
sensitively on the assumed line width. The most reasonable solution,
based on the \formalde\ and CS lines and the \eighteenco\ map, is to
simultaneously fit the peak intensity with three Gaussians, centered
on \vlsr= 2.7, 3.8, and 5.0~\kms, with FWHM=0.5, 0.8, and
1.5~\kms. When needed, we allow a 30\% variation in the FWHM, and
0.20~\kms\ in \vlsr.  Some of these small variations may be real, but
given the complexity of the line profiles and spatial distribution of
the various components, we will not seek a physical interpretation for
this.  The integrated intensities are summarized in
Table~\ref{t:decomp}, and the physical conditions are derived from
these in \S 4.

We conclude that when studying individual protostars in the $\rho$ Oph
cloud, it must be realized that many different physical components are
present within single dish beams of $\sim15''$ or larger.  The spatial
information of interferometers is essential here.

\begin{figure*}[p]
\center
\includegraphics[angle=90, scale=0.45]{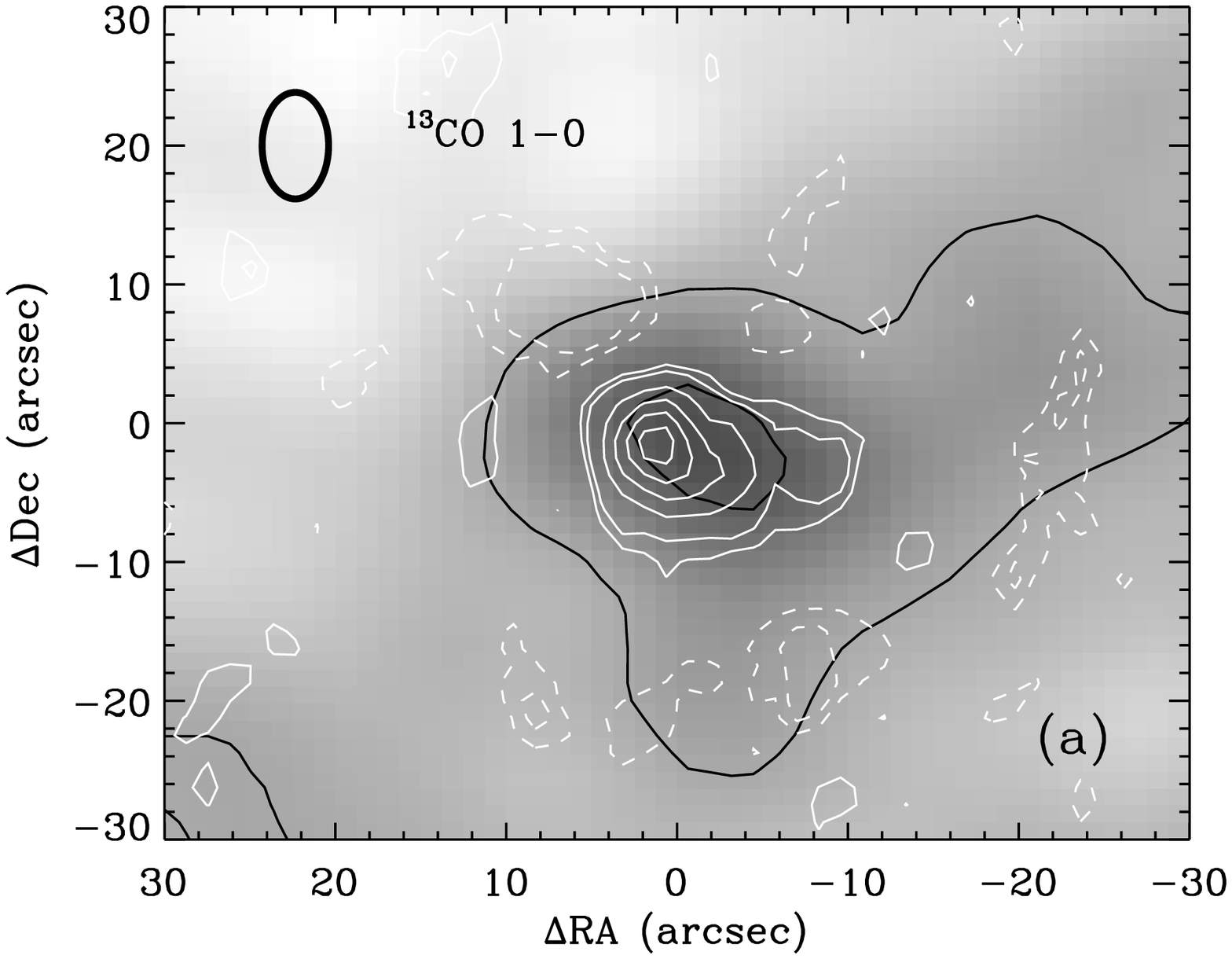}
\includegraphics[angle=90, scale=0.45]{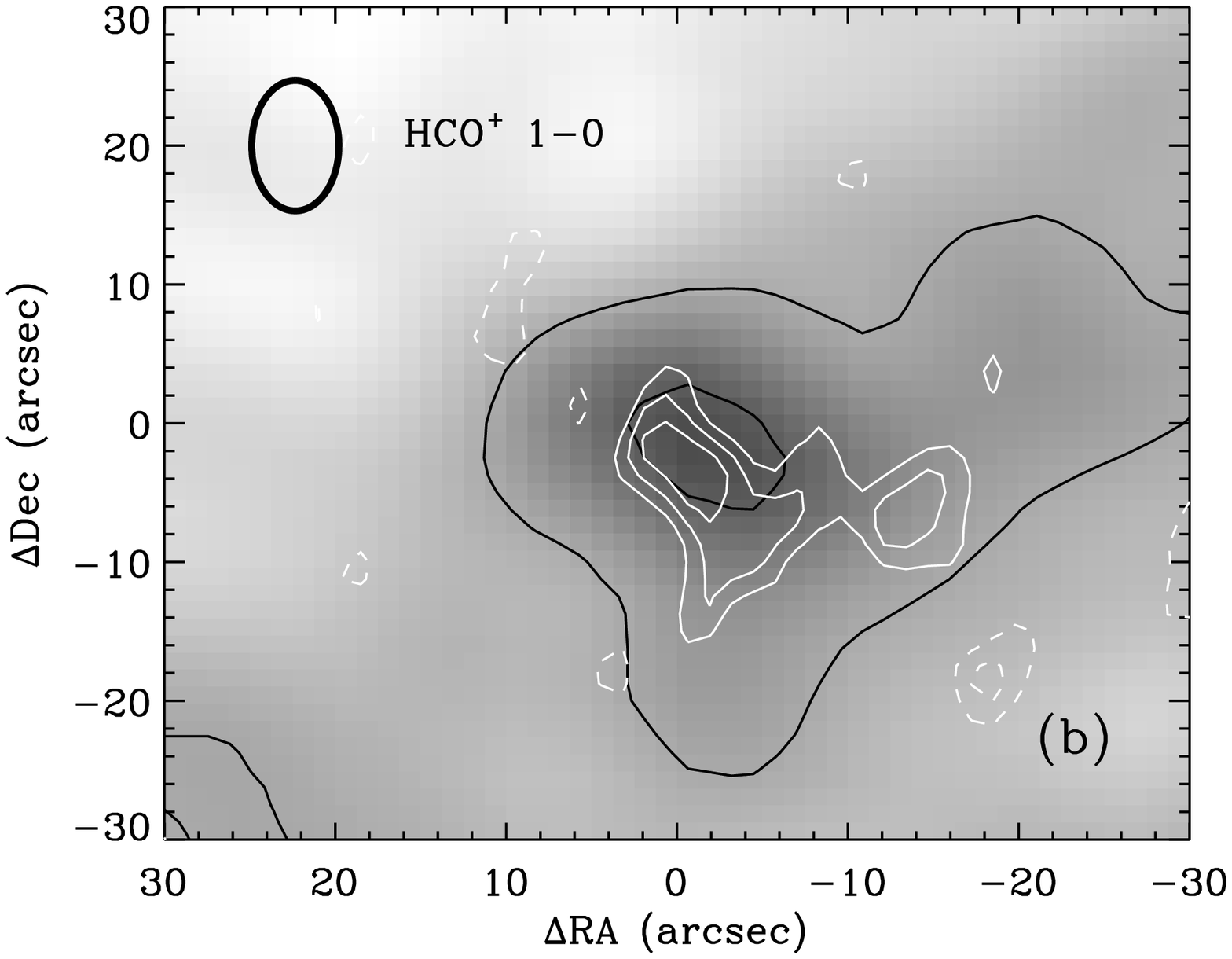} 
\caption{Integrated OVRO \thirteenco\ 1-0 (a) and \hcop\ 1-0 (b)
emission in the Elias 29 region (white contours), superposed on the
1.3 mm continuum map (gray scale and black contours).  Contour levels
are drawn at -3,-2, 2, 3, 6, ...,15 and -3,-2, 2, 3, and 4 times the
noise level (0.48 and 0.90 Jy/beam for \thirteenco\ 1-0 and \hcop\ 1-0
resp.). The contours for negative signals are given by white dashed
lines.  Contour levels for the 1.3 mm map are the same as in
Fig.~\ref{f:hcop}. The synthesized beam size is given in the top left
corners.}~\label{f:ovromap}



\center
\includegraphics[angle=90, scale=0.45]{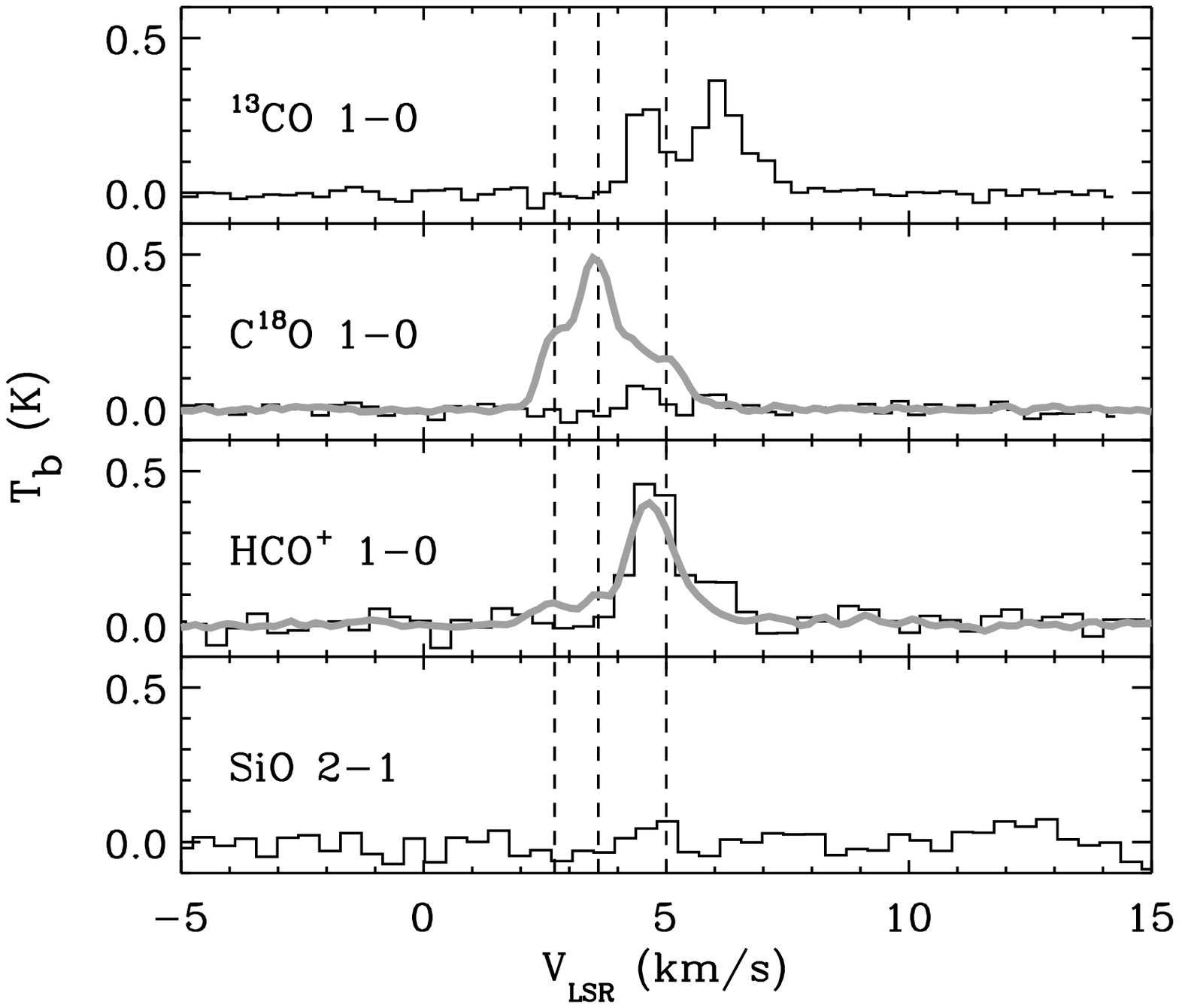} 
\caption{Average line profiles of the central region with more than
2$\sigma$ emission in the OVRO maps, which is an area of 380
arcsec$^2$ for \thirteenco\ and 650 arcsec$^2$ for \hcop. For the
non-detected \eighteenco\ and SiO lines we took the central 400
arcsec$^2$.  The brightness temperature is correspondingly diluted by
factors of 7.4, 5.9, 5.8, and 6.4 from top to bottom respectively to
derive the expected signal in the single dish NRAO-12m beam.  The
smooth gray line for C$^{18}$O and HCO$^+$ is the NRAO-12m single dish
emission, divided by factors of 13 and 9 respectively (for
clarity). The vertical dashed lines represent the single dish emission
line components shown in Fig.~\ref{f:sdish}.}~\label{f:ovrospec}

\center
\includegraphics[angle=90, scale=0.44]{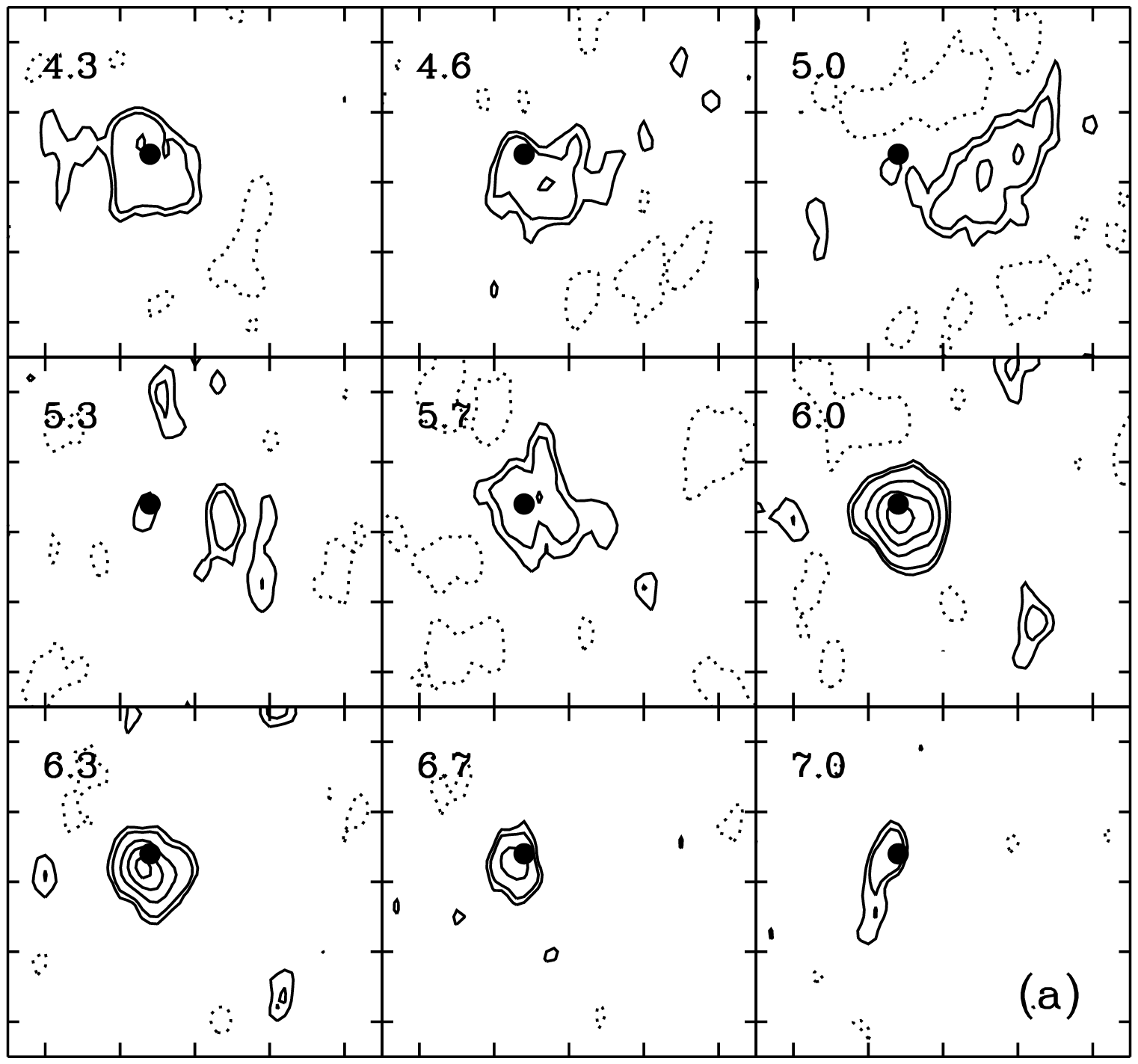}
\includegraphics[angle=90, scale=0.44]{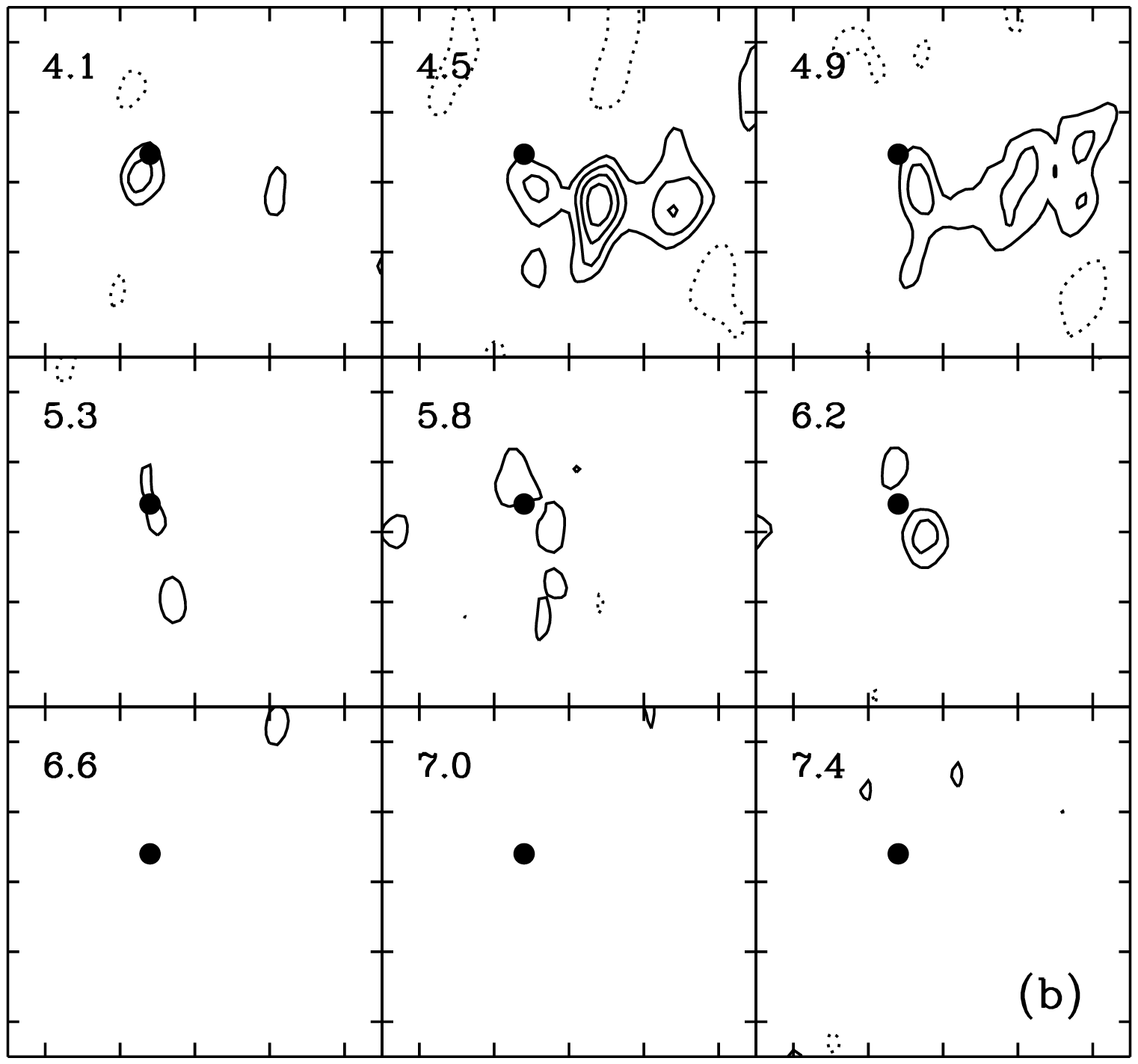} 
\caption{Channel maps of OVRO \thirteenco\ 1-0 (a) and \hcop\ 1-0 (b)
emission in the Elias 29 region. The map size is 50$''$.  The bullets
indicate the map centers, and facilitate comparison of the
panels. Contour levels are drawn at -2, 2, 3, 6, 9, 12 (\thirteenco)
and -2, 2, 3, 4, and 5 times the noise level (0.13 and 0.26 Jy/beam
resp.), with the negative contour indicated with dots. Extended
emission toward the west is present at velocities near 5.0 \kms\ in
both lines and coincides with the large scale dense ridge seen in our
\hcop\ 3-2 single dish map.}~\label{f:ovrochan}
\end{figure*}

\subsection{Small Scale Emission: Interferometer Maps}

In the OVRO interferometer maps, bright \thirteenco\ and \hcop\ 1-0
emission is detected in the direct neighbourhood of Elias 29
(Fig.~\ref{f:ovromap}). The \thirteenco\ emission is most strongly
peaked on the infrared position, and with a spatial FWHM$\sim 6''$
(900 AU) it is resolved along the EW direction, where the OVRO beam is
smallest (3$''$). This is the Elias 29 core, which likely consists of
a disk/envelope system (\S 4.2). An extension of $\sim 12''$ toward
the southwest is visible in both lines, but most prominently in \hcop\
1-0, which must be attributed to the ridge discussed above. Indeed, on
this small scale the \hcop\ emission is also parallel to the IRAM-30m
1.3 mm continuum emission (Fig.~\ref{f:ovromap}).

\begin{figure*}[b!]
\center
\includegraphics[angle=270, scale=1.10]{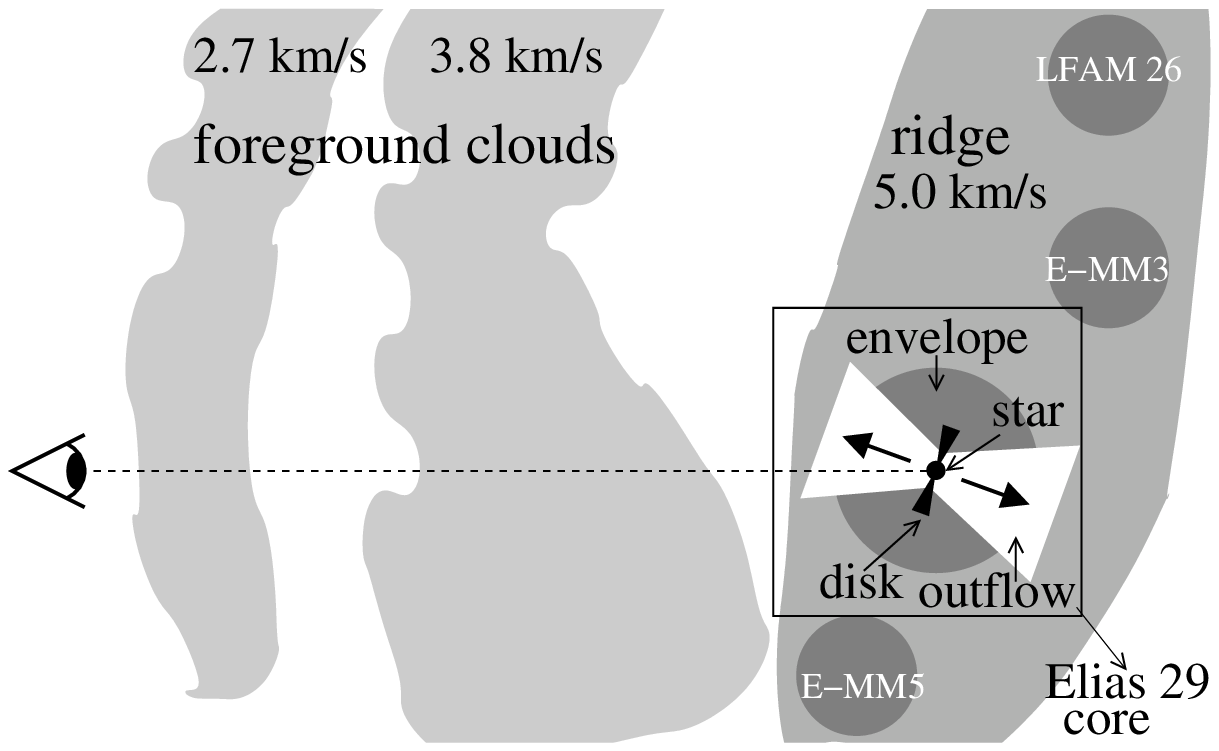}
\caption{Schematic overview of the proposed geometry of Elias~29 and
its environment (not to scale). The line of sight to Elias~29 may
cross part of the envelope if the system is slightly more inclined
than shown here. Most of the ice-absorption bands originate in the
foreground clouds.}\label{f:schem}
\end{figure*}

The OVRO \thirteenco\ and \hcop\ 1-0 spectra peak between \vlsr=5--7
\kms, and show no evidence for the strong, extended 2.7 and 3.8 \kms\
foreground emission seen in the single dish maps
(Fig. \ref{f:ovrospec}).  The \eighteenco\ 1-0 emission is {\it
spectacularly} absent in the interferometer spectrum; it is at least a
factor of 10 weaker compared to the single dish data. After dilution
to the NRAO-12m beam (taking the emitting area from the OVRO
\thirteenco\ image), the OVRO \eighteenco\ signal is even a factor of
60 weaker than the bright detection with the NRAO-12m telescope. This
shows again that the bulk of the molecular material in the Elias 29
line of sight is not associated with Elias 29, but instead is present
in extended foreground clouds, which are resolved out by the OVRO
interferometer.

It is interesting to note that the OVRO line profiles are double
peaked (Fig. \ref{f:ovrospec}), with a 5.0 \kms\ component emitting in
the dense ridge (Fig. \ref{f:ovrochan}). Emission at the central
infrared position is strongest at 6 \kms. Whether this is a true
dynamical difference between the Elias 29 core and the ridge, or
artificially created by self-absorption at $\sim 5.5$ \kms, is
difficult to answer at present. In this paper we will keep referring
to this as the 5.0 \kms\ component.  A disk/envelope/ridge
decomposition is attempted in \S 4.2.

\section{Discussion: Physical Conditions}

It is our aim to derive the physical structure of the surroundings of
Elias~29, and to locate the origin of the ices, seen abundantly along
this line of sight \citep{boog00b}. In \S 4.1 the extended clouds at
2.7 and 3.8 km~s$^{-1}$ identified above are discussed; \S\S 4.2 and
4.3 describe the more immediate circumstellar environment of Elias~29
in terms of a near face-on disk, a remnant envelope, and the dense
ridge from which the star may have formed (see
Fig.~\ref{f:schem}). The physical conditions in each of the components
are constrained using the intensities and intensity ratios of the
single dish and interferometer line emission, the 1.3~mm continuum
emission \citep{mott98}, and the infrared SED \citep{boog00b}.  This
is linked to the ice observations in \S 4.4.

\subsection{Foreground Clouds}

The C$^{18}$O line emission at 2.7 and 3.8 km~s$^{-1}$ is extended
over several arcminutes (Fig.~\ref{f:csochan}), and thus is associated
with the overall $\rho$~Oph cloud complex.  Assuming that the clouds
are sufficiently homogeneous, we can use the ratios of the decomposed
line intensities (Table~\ref{t:decomp}) to find the density and
temperature, and subsequently the column density, in each of the
clouds. We use the escape-probability method described by
\citet{jans95} to calculate the molecular excitation.

The most useful constraints on the density and temperature are given
by the intensity ratios of \eighteenco\ 1--0/3--2 and \hcop\
1--0/3--2.  Fig.~\ref{f:phys} visualizes the density and temperature
values allowed by the observed ratios, taking into account
line-opacity effects. The 2.7 \kms\ component has a temperature
$T=15\pm 5$ K, and a density $n$(H$_2$)=($1\pm0.5$) $\times 10^4$
\cubcm.  For the 3.8 \kms\ component, the density is less than $10^5$
\cubcm, but the temperature cannot be significantly constrained from
the line ratios.  Here we use the far infrared SED to find that
$T_{\rm kin} = 25 \pm 15$ K (\S 4.2). Taking a C$^{18}$O abundance
with respect to H$_2$ of $3.6\times 10^{-7}$
($N$[CO]/$N$[C$^{18}$O]=560, \citealt{wils94}; $N$[H$_2$]/N[CO]=5000,
\citealt{lacy94}), the H$_2$ column densities of the 2.7 and 3.8
km~s$^{-1}$ clouds are $5\times 10^{21}$ cm$^{-2}$ and $1.4\times
10^{22}$ cm$^{-2}$, respectively (Table~\ref{t:phys}).  This also
assumes that CO is not strongly depleted in these clouds, which is
validated in \S 4.5.  These parameters fit the optically thin
C$^{17}$O lines also, indicating that a correct line opacity was used

With these physical parameters at hand, we calculate that the optical
depth of the clouds in the CO 5-6 transition is $\sim$4. This is
sufficiently high to explain the absorption in the CO 6-5 lines at 2.7
and 3.8 \kms\ (Fig.~\ref{f:sdish}). The bright CO 6-5 emission is
closely associated with Elias 29 (C. Ceccarelli et al., in prep.), and
therefore the 2.7 and 3.8 \kms\ clouds are located in front of this
source.

Independent information on the physical conditions in these extended
foreground clouds is obtained from the SED longward of 100 \mum\ (\S
4.2).  The columns and temperatures compare also favorably with the
extended cold dust found by sensitive 180--1100 $\mu$m balloon-based
measurements (\citealt{rist99}; Table~\ref{t:colden}).

The observed lines of CS, H$_2$CO, and CH$_3$OH do not further
constrain the physical conditions, but they can be used to derive
abundances.  These are listed in Table~\ref{t:phys}, and further
discussed in \S 4.5.

\vbox{
\begin{center}
\leavevmode 
\hbox{%
\epsfxsize\hsize
\includegraphics[angle=270, scale=0.45]{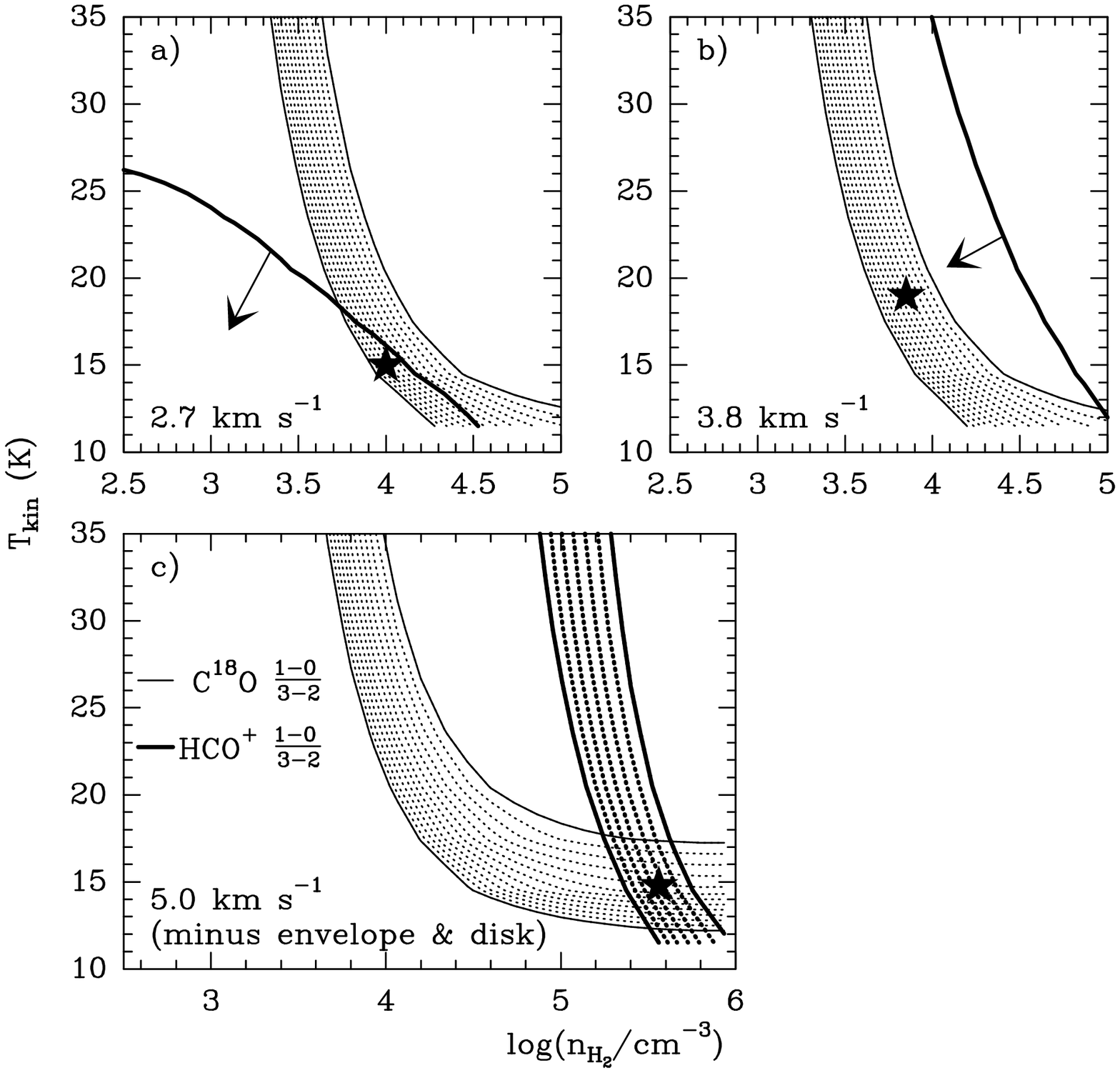}}
\figcaption{Temperature and density range constrained by intensity ratios
of \eighteenco\ 1--0/3--2 (thin lines) and HCO$^+$ 1--0/3--2 (thick
lines) for the foreground clouds at \vlsr =2.7 ({\bf a}) and 3.8 \kms\
({\bf b}), and for the ridge material at 5.0 \kms\ ({\bf c}). The
intensity ratios for the ridge material have been corrected for the
expected contribution from the envelope and disk of Elias~29. The
filled star indicates the adopted temperature and density for each of
the components.\label{f:phys}}
\end{center}}


\subsection{The Circumstellar Environment of Elias 29}

\begin{figure*}[t!]
\center
\includegraphics[angle=270, scale=0.60]{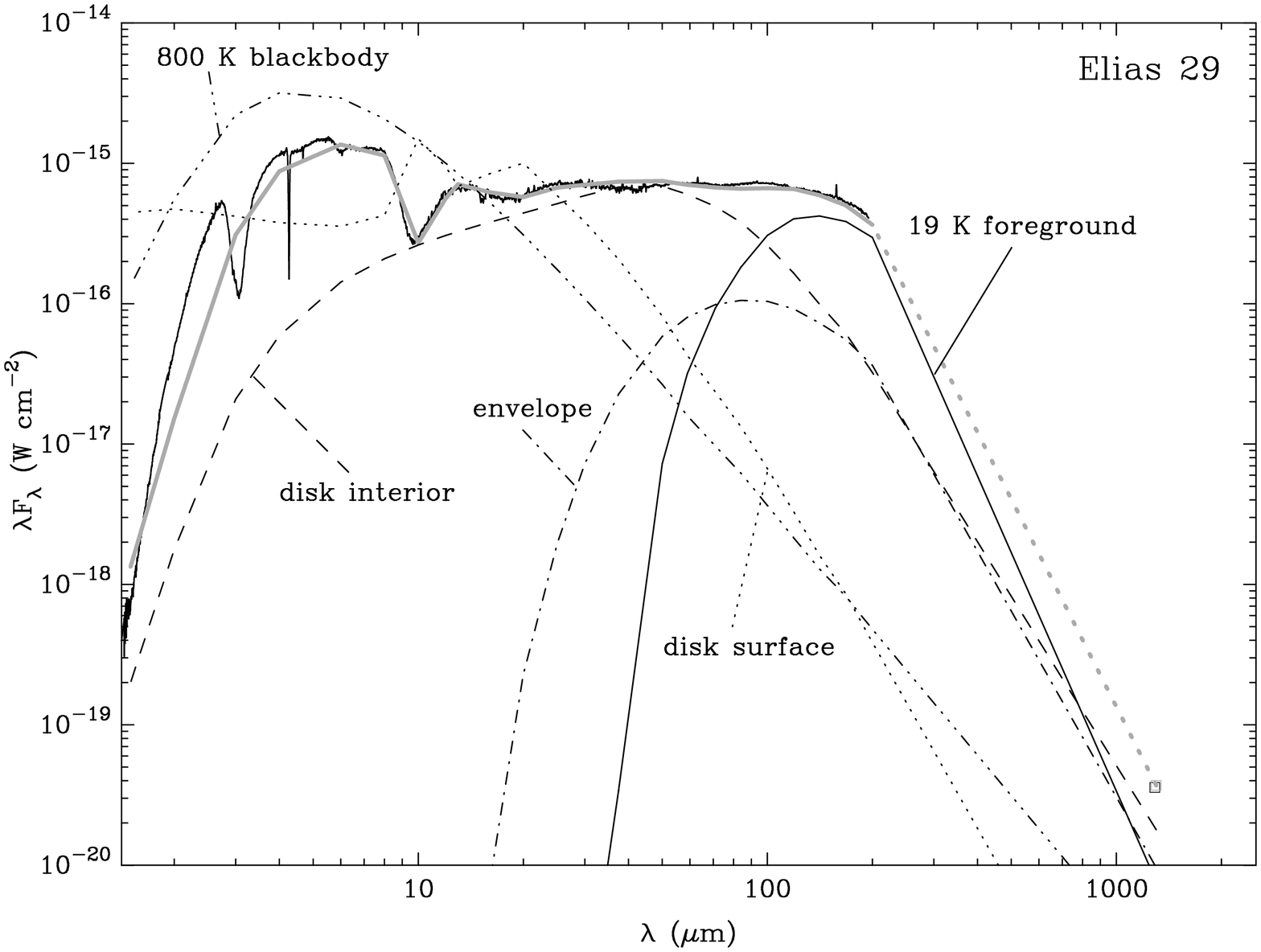}
\caption{Infrared SED of Elias~29 as observed by ISO LWS and SWS (thin
line with ice and silicate absorption features; \citealt{boog00b})
combined with the 1.3~mm flux in the 15\arcsec\ IRAM-30m beam (open
square; \citealt{mott98}). The thick gray line shows the modeled SED
as described in \S 4.2.  It consists of four components: (1) a
0.012~M$_\odot$ flared circumstellar disk following \citet{chia97}
with a two-layer vertical temperature structure (cool interior
[long-dashed line] and superheated surface [dotted line]); (2) a
0.12~M$_\odot$ collapsing envelope (dash-dot line); (3) an {\it ad
hoc} 800~K central blackbody with a 1.4~AU radius (dash-triple dot
line), required to fit the flux maximum at $\lambda\sim 5$ \mum\ and
probably corresponding to the warmest regions of the disk close to the
star and/or the stellar photosphere extincted by the disk; and (4) a
19~K foreground layer (thin solid line peaking at $\sim$120 \mum) with
a column density of $N({\rm H_2}) = 1.9\times 10^{22}$ cm$^{-2}$ as
derived for the material at \vlsr =2.7 and 3.8 \kms\ in \S 4.1, which
is responsible for the continuum absorption below 3 \mum\ in the final
four component model, and for the prominent silicate and ice
absorption bands.}\label{f:sed}
\end{figure*}

Understanding the circumstellar environment of Elias~29 requires the
combination of several pieces of crucial information: the single-dish
\hcop\ 3--2 map and 1.3~mm dust-continuum distribution
(Fig.~\ref{f:hcop}), the spatially resolved emission in $^{13}$CO 1--0
and \hcop\ 1--0 observed by OVRO (Fig.~\ref{f:ovromap}), and the
infrared SED obtained by the SWS and LWS spectrometers of the ISO
satellite (\citealt{boog00b}; Fig.~\ref{f:sed}).

The \hcop\ 3--2 and 1.3~mm continuum maps show that Elias~29 is
located in a narrow, dense ridge, $\lesssim 30''$ wide and several
arcminutes long. The strongest \hcop\ 3--2 peak in the ridge is
situated near Elias 29.  The OVRO images and channel maps
(Figs.~\ref{f:ovromap} and~\ref{f:ovrochan}) show that part of the
$^{13}$CO and \hcop\ emission is centered on Elias~29, while a
significant fraction follows the crest of the ridge, $10''$--$20''$
offset to the west/southwest.  This shows that multiple components are
present in the immediate vicinity of Elias~29 (i.e., within a typical
single-dish beam), in addition to the multiple velocity components
along the line of sight (\S 4.1). The velocity coincidence between the
ridge and Elias~29 further supports that the star formed from the
ridge.

\subsubsection{SED and Continuum Modeling}

The most important clues about the nature of Elias~29 are offered by
the infrared SED. Because of the large beam with which these
measurements were taken (\S 2), the SED reflects not only Elias~29 but
also many of the other components identified above. Our ability to
derive accurate parameters for each component therefore proved
essential in helping to understand the nature of Elias~29. The SED
(Fig.~\ref{f:sed}) is remarkably flat between 10 and 200 $\mu$m. At 10
$\mu$m there is a prominent silicate absorption band, and at 5 $\mu$m
the SED shows a broad peak on top of which a number of sharp ice
absorption features are present. Below 2 $\mu$m the emission drops off
sharply. Flat protostellar SEDs are generally explained by a
circumstellar disk. Compared to other circumstellar material
distributions, a disk, especially a flared one, has the surface-area
vs. temperature distribution required to create a flat SED.

For our analysis, we adopt the flared disk model of \citet{chia97}
with the parameters listed in Table~\ref{t:model}.  The flatness of
the SED is explained in this model by a superheated surface layer. The
continuum emission of the disk is modeled using the radiative transfer
part of the Monte-Carlo code of \citet{hoge00}; this part of the code
simply calculates the expected continuum emission from a given
distribution of density and temperature by building up a grid of
points on the sky, for each of which the radiative transfer is solved
along straight lines. The resulting grid is then convolved with the
appropriate beam sizes.  Because of the large number of parameters
involved, their values should be considered as representative only. We
believe that the general characteristics of the model are firmly
established, however.  The flatness of the SED over a large wavelength
range (4-100 \mum; Fig.~\ref{f:sed}) limits the inclination of the
system to $<60^\circ$ ($90^\circ$ being edge-on; \citealt{chia99}).
Such a low inclination cannot explain the deep ice and silicate
absorption features that are visible in the observed SED; these must
originate in the envelope and foreground clouds.  The low disk
inclination is compatible with the very high velocity ($\sim 80$
\kms), variable outflowing hot CO gas seen in this object
(A.C.A. Boogert, G.A. Blake, \& M.R. Hogerheijde, in prep.). A fully
face on orientation is however not likely, given the presence of low
surface brightness scattered K band light out to distances of 15$''$
(2400 AU) from the central object (Zinnecker, Perrier, \& Chelli
1988).  This light presumably traces the outflow lobes, adjacent to
the remnant envelope and the outer edges of the disk.  The emission is
extended along the southwest/northeast direction, perpendicular to the
high velocity outflowing gas (C. Ceccarelli et al., in prep.), with an
axis ratio compatible with an inclination of roughly $\rm \sim
30^o$. Flattening on a similar scale is observed in the OVRO images
and in the 1.3 mm continuum IRAM-30m single dish map.

\begin{table*}[p]
\center
{\footnotesize
\caption{Physical Conditions in Elias 29 Foreground Clouds and Ridge, and
Comparison with Other Clouds}~\label{t:phys}
\center
\small
\begin{tabular}{llllllll}
\tableline
\noalign{\smallskip} 
Quantity & Unit      & \multicolumn{3}{c}{Cloud component} & TMC 1$^d$ & \multicolumn{2}{c}{NGC 1333$^e$}\\
\noalign{\smallskip}  
\cline{3-5}
\cline{7-8}
\noalign{\smallskip}  
         &       & 2.7~\kms    & 3.8~\kms & 5.0~\kms &    & IRAS 4A & IRAS 4B \\
\noalign{\smallskip}  
\tableline
\noalign{\smallskip} 
$b_{\rm D}$      & [\kms]           & 0.3       & 0.5       & 1.0      &\nodata  & \nodata & \nodata\\
$T_{\rm kin}$    & [K]              & 15 (5)    & 25 (15)   & 15 (5)    &\nodata  & \nodata & \nodata\\
$n{\rm (H_2)}$   & [10$^4$ \cubcm]  & 1.0 (0.5) & $<10$     & 40 (20)   &\nodata  & \nodata & \nodata\\
$N{\rm (H_2)}$   & [10$^{22}$ \sqcm] & 0.5 (0.2) & 1.4 (0.2) & 1.0 (0.2) & 1.0     & 14      & 6 \\ 
$X$(\eighteenco)$^a$& [10$^{-9}$]   & 360       & 360       & 360       & 304     & 40       & 70  \\
$X$(C$^{17}$O)$^a$& [10$^{-9}$]     & 110       & 110       & 110       & 95      & 7.6     & 19  \\
$X$(CS)          & [10$^{-9}$]      & 6.0       & 15.0      & \nodata$^b$ & 6     & 1.2     & 0.2  \\
$X$(\hcop)        & [10$^{-9}$]     & 0.6       & 0.7       & \nodata$^b$ & 9     & 0.4     & 0.1 \\
$X$(ortho \formalde)& [10$^{-9}$]   & 3.0       & 8.0       & \nodata$^b$ & 50    & 0.4     & 0.9  \\
$X$(para \formalde)& [10$^{-9}$]    & $<$1.0    & 2.7       & \nodata$^b$ &\nodata  & \nodata & \nodata\\ 
$X$(\methanol)       & [10$^{-9}$]  & 0.4       & 0.5       & \nodata$^b$ & 3     & \nodata & \nodata\\
$A_{\rm V}$      & [mag]            & 2.9       & 8.2       & $<5.9^c$   &\nodata & \nodata & \nodata\\
\noalign{\smallskip} 
\tableline
\multicolumn{8}{p{15cm}}{General: all abundances $X$ are w.r.t. H$_2$ and have an error of 50\%}\\
\multicolumn{8}{p{15cm}}{$^{\rm a}$ From H$_2$ using standard relations (see text), except TMC-1}\\
\multicolumn{8}{p{15cm}}{$^{\rm b}$ Line intensities in ridge have to
be corrected for expected contribution from envelope and disk (see
text). No independent determination of abundances is therefore possible.}\\
\multicolumn{8}{p{15cm}}{$^{\rm c}$ Upper limit; only a fraction of
the ridge material may actually be in front of Elias~29.}\\
\multicolumn{8}{p{15cm}}{$^{\rm d}$ on `CP' peak; \citealt{prat97, ohis98}}\\
\multicolumn{8}{p{15cm}}{$^{\rm e}$ \citealt{blak95}, assuming
$N({\rm H_2})$ from dust}\\ \noalign{\smallskip}
\end{tabular}}

\center
{\footnotesize
\caption{Column Density of Cold Gas Derived by Different Methods}~\label{t:colden}
\center
\begin{tabular}{llll}
\tableline
\noalign{\smallskip} 
Method & Component & $N{\rm (H_2)}$$\rm ^a$  & $A_{\rm V}$$\rm ^b$\\
       &           & 10$^{22} $\sqcm         &                    \\
\noalign{\smallskip}  
\tableline
\noalign{\smallskip} 
silicate absorption$\rm ^c$ &                   & 2.5-6                & 14-34  \\
4.7 \mum\ \thirteenco\ absorption$\rm ^d$ & \bdop=1.0 \kms & 3 (0.5)   & 17 (3) \\
                        & \bdop=0.7 \kms        & 11 (4)               & 63 (23)\\
\eighteenco\ 2-1 emission& \vlsr=2.7    & 0.5 (0.2)            & 2.9 (1.1)\\
                        & \vlsr=3.8     & 1.4 (0.2)            & 8.0 (1.1)\\
                        & \vlsr=5.0     & 1.0 (0.2)            & 5.7 (1.1)\\
ISO SED and IRAM-30m 1.3 mm & envelope              & 0.3-1                & 1.7-6  \\
0.18--1.1 mm continuum$\rm ^e$ & 10--15 K       & 0.5--5               & 2.9-29 \\
X rays$\rm ^f$          &                       & 3.8                  & 22     \\
\noalign{\smallskip} 
\tableline
\multicolumn{4}{p{12cm}}{$\rm ^a$ assuming $N{\rm (H_2)}/N{\rm (CO)}=5\times 10^3$;
$\rm ^b$ conversion factor $A_{\rm V}=N{\rm (H_2)}\times 8.6/15\times
10^{21}$ \citealt{bohl78};
$\rm ^c$ see \citealt{boog00b}; 
$\rm ^d$ A.C.A. Boogert et al., in preparation; 
$\rm ^e$ Balloon experiment: \citealt{rist99};
$\rm ^f$ \citealt{iman01}}\\ 
\end{tabular}}
\end{table*}

The flux between 20 and 40 $\mu$m indicates a temperature scaled
upward by a factor of 2.25 with respect to the values used by
\citet{chia97}. The central star is therefore approximately 50 times
brighter than their standard model, presumably because it has a higher
mass (using the scaling relation $T\propto L^{1/5}$;
\citealt{chan00}).  The disk midplane dominates the emission beyond 25
$\mu$m, and the shape of the SED at these wavelengths limits the disk
radius to $\sim$ 500 AU.  The warmer surface layer dominates the
shorter wavelengths and generate a silicate {\sl emission} feature at
10 $\mu$m (Fig.~\ref{f:sed}). The disk model can explain only
$\sim$20\% of the emission maximum around 5 $\mu$m. Indeed, speckle
observations of Elias 29 reveal the presence of a $\sim$400 AU radius
thermally emitting region ($T$ a few 100 K; i.e. not scattered light)
responsible for $\sim$20\% of the M band flux \citep{zinn88}, which,
in our model, is explained by the warm disk surface layer.  The
observed remaining 80\% of the M band emission originates from hot
dust within a few AU from the central object.  Perhaps this is a
puffed up hot rim at the inner disk edge, just outside the region
where dust has evaporated ($T_{\rm subl}\sim 1500$ K;
\citealt{dull01}; \citealt{zinn88}).  The innermost, hottest, part of
the rim may be related to the 0.25 AU radius structure found in lunar
occultation K band observations \citep{simo87}.  We do not attempt to
fit the inner regions and stellar photosphere with that level of
detail, but instead use an 800~K blackbody with an effective radius of
1.4~AU to explain the remaining 80\% of M band emission, and roughly
the shape of the SED at shorter wavelengths (Fig.~\ref{f:sed}).

The disk alone is also insufficient to explain the emission beyond
$\sim 55$ $\mu$m. The single-dish line emission clearly reveals the
presence of appreciable columns of material at 2.7 and 3.8 km~s$^{-1}$
(\S 4.1). When filling the ISO beam with these columns, the emission
between 55 and 180 $\mu$m can be fit at a dust temperature of 19~K.
This material provides sufficient opacity in the 10 \mum\ silicate
band to turn the disk's emission feature into the observed absorption
feature. The opacity shortward of 2 $\mu$m is also sufficient to
explain the observed steep drop in emission at short wavelengths.
This strongly suggests that a large fraction of the 2.7 and 3.8
km~s$^{-1}$ material is located in front of Elias~29, as was found
from the absorption in the \twelveco\ emission lines as well (\S 4.1).

\begin{table*}[b!]
\center
{\footnotesize
\caption{Parameters of Disk and Envelope Model}~\label{t:model}
\begin{tabular}{lp{10cm}}
\tableline
\noalign{\smallskip} 
Item & Value \\
\noalign{\smallskip}  
\tableline
\noalign{\smallskip} 
\multicolumn{2}{c}{Disk: \citealt{chia97}}\\
\noalign{\smallskip}  
\tableline
\noalign{\smallskip} 
Temperature & $2.25\times$ standard model of Chiang \& Goldreich \\
Dust opacity & \citealt{osse94}, MRN model with $10^6$
  years of coagulation at $10^5$ cm$^{-3}$  \\
Inner radius & 0.01 AU \\
Outer radius & 500 AU \\
Mass & 0.012 M$_\odot$ \\
\% mass in superheated surface layer & 2.5\% \\
\noalign{\smallskip}  
\tableline
\noalign{\smallskip} 
\multicolumn{2}{c}{Envelope: \citealt{shu77}}\\
\noalign{\smallskip}  
\tableline
\noalign{\smallskip} 
Sound speed & 0.13 km~s$^{-1}$ \\
Age & $2.2\times 10^5$ yr \\
Radius of collapse expansion wave & 6000 AU \\
Temperature & $35\,(r/1000\,{\rm AU})^{-0.4}$ K \\
Dust opacity & \citealt{osse94}, MRN model with $10^6$
  years of coagulation at $10^5$ cm$^{-3}$ \\
Inner radius & 500 AU (3\arcsec) \\
Outer radius & 6000 AU (39\arcsec)\\
Mass & 0.12 M$_\odot$ \\
\noalign{\smallskip} 
\tableline
\end{tabular}}
\end{table*}

\vbox{
\begin{center}
\leavevmode 
\hbox{%
\epsfxsize\hsize
\includegraphics[angle=270, scale=0.35]{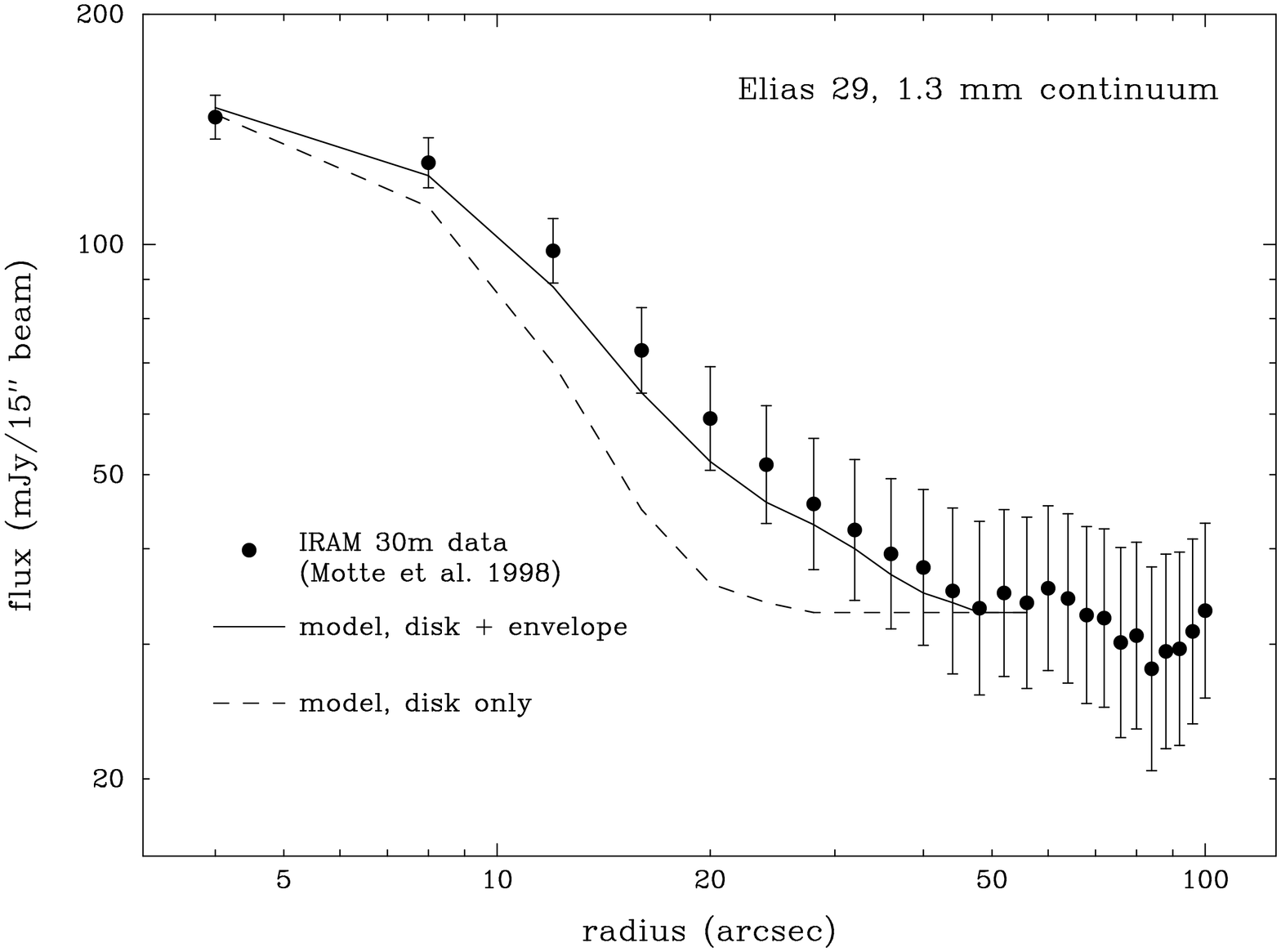}}
\figcaption{Radial intensity profile along the ridge centered on Elias~29
obtained from the 1.3~mm map of \citet{mott98} (bullets with error
bars), compared to the predicted profile based on a model for Elias~29
with a disk and an envelope (solid line) and with a disk only (dashed
line). Both models include a 19~K, $N({\rm H_2}) = 6\times 10^{21}$
cm$^{-2}$ background to fit the levelling off of the emission beyond a
radius of $25-50''$. Note that the true total background column,
filtered out of the IRAM-30m 1.3 mm data, is a factor of 3 higher (\S
4.2.1).\label{f:13mmprof}}
\end{center}}

While the model of disk and foreground clouds fits the SED well, it
produces a 1.3 mm continuum emission that is a factor of 3 larger than
observed with the IRAM-30m telescope, at large distances from the
object ($\sim 30$ mJy; Fig.~\ref{f:13mmprof}). This may result
directly from the adopted 'dual-beam' mapping procedure of the
IRAM-30m observations, which filters out emission on the chopping
scale \citep{mott98, john00a}. Additionally, the model of disk and
foreground predicts a source size at 1.3~mm that is too small
(Fig.~\ref{f:13mmprof}).  The observed profile indicates that the
source extends to a radius of at least 3800~AU (25\arcsec), much
larger than is realistic for a circumstellar disk.  Instead, it
suggests that Elias~29 and its disk are embedded in a residual cloud
condensation from which the system formed.  For lack of stronger
constraints, we model this envelope (as we will refer to this
condensation from now on) with the inside-out collapse model of
\citet{shu77}. Earlier work (e.g., \citealt{hoge00_s};
\citealt{cecc00}) has shown that this model provides an adequate
description of the density and temperature of envelopes around
YSOs. The two parameters of this model are the sound speed and the
time since the collapse started, for which we take, rather
arbitrarily, 0.13 \kms\ and $1.2\times 10^5$ yr respectively. This
gives an envelope mass of 0.12 $M_\odot$ and a radius of the `collapse
expansion wave' of 6000 AU. Additional parameters are listed in
Table~\ref{t:model}.  This model, combined with the disk, produces a
radial emission profile that fits the measured 1.3~mm continuum
distribution much better than the disk alone (Fig.~\ref{f:13mmprof}).
The SED below $\sim 100$ \mum\ is remarkably little affected by the
envelope, even though with 0.12~M$_\odot$ it is much more massive than
the disk (0.012~M$_\odot$; Table~\ref{t:model}). The material in the
disk, of course, has a somewhat higher temperature range than the
envelope (200--33 K vs. 44--17 K, respectively), yielding
significantly more infrared flux. The envelope produces only a small
amount of additional absorption in the 10 $\mu$m silicate band.

The parameters for the disk and envelope derived above
(Table~\ref{t:model}) maximize the contribution to the SED from the
disk and minimize the mass of the envelope. There are several pieces
of evidence that favor, but not proof, this maximum disk model above
models with smaller disks. First, as for sources with directly
detected disks \citep{hoge98}, the disk contributes about half of the
1.3 mm continuum flux ($\sim$ 120 mJy/15\arcsec\ beam; foreground
subtracted).  Indeed, assuming that the weak 3 mm continuum flux
within the small OVRO beam (\S 2.2) orginates from the disk, then the
expected 1.3 mm disk flux would be half the single dish flux ($F$[1.3
mm]=49$\pm$14 mJy, taking spectral index 2.5; \citealt{hoge98}).
Second, the abovementioned 400 AU radius structure, emitting thermally
at 5 \mum, may well be the surface of a large, warm disk.  And third,
a large face-on flared disk is the simplest way to explain the
observed flat infrared SED.

Nevertheless, the evidence for a large 500 AU radius disk is not
conclusive and therefore we also try to maximize the envelope's
contribution in our model.  We find that its mass can be as large as
0.33 $M_\odot$, while the corresponding disk mass needs to be reduced
to 0.002 M$_\odot$, in part by reducing the disk's radius to 30 AU to
preferentially decrease the amount of cool material. The disk's
temperature scaling does not change appreciably with respect to that
used before. It must be emphasized that in this minimum disk scenario,
the flatness of the infrared SED is the result of a ``disk/envelope
conspiracy'', rather than a direct consequence of the flared disk's
properties. Although the disk and envelope masses are different from
those listed in Table~\ref{t:model}, they are comparable to within the
same order of magnitude and this level of uncertainty does not affect
our conclusions. In the remainder of this section we will adopt the
slightly preferred values from Table~\ref{t:model}, i.e. the maximum
disk case.

\subsubsection{Emission Line Modeling}

\begin{table*}
\center
{\footnotesize
\caption{Predicted Line Intensities}~\label{t:predict}
\begin{tabular}{lrcccc}
\tableline
\noalign{\smallskip} 
Line & Observed$\rm ^a$ & Envelope+Disk$\rm ^b$ & Envelope$\rm ^c$ & Disk$\rm ^d$ & Adopted abundance\\
\noalign{\smallskip} 
\tableline
\noalign{\smallskip} 
\multicolumn{6}{c}{Single dish observations (K~km~s$^{-1}$)}\\
\noalign{\smallskip} 
\tableline
\noalign{\smallskip} 
C$^{18}$O 1--0 & 3.97  & 0.57 & 0.52 & 0.05 & $3.6\times 10^{-7}$\\
C$^{18}$O 2--1 & 9.50  & 3.22 & 2.64 & 0.75 & \\
C$^{18}$O 3--2 & 6.78  & 4.57 & 2.97 & 2.15 & \\
HCO$^+$ 1--0 & 4.26    & 1.01 & 0.99 & 0.10 & $3\times 10^{-9}$\\
HCO$^+$ 3--2 & 3.17    & 2.32 & 1.98 & 3.43 & \\
HCO$^+$ 4--3 & 0.76    & 1.51 & 0.69 & 2.87 & \\
H$^{13}$CO$^+$ 1--0 & 0.33  & 0.04 & 0.03 & 0.01 & $4\times 10^{-11}$ \\
CS 2--1 & 1.72  & 0.78 & 0.76 & 0.13 & $1\times 10^{-8}$\\
CS 5--4 & 0.82  & 1.49 & 0.40 & 2.14 & \\
CS 7--6 & $<0.13$ & 4.44 & 0.11 & 4.39&  \\
C$^{34}$S 2--1 & $<0.05$ & 0.07 & 0.06 & 0.02 & $4\times 10^{-10}$\\
ortho-H$_2$CO $2_{12}$--$1_{11}$ & 1.03 & 1.16 & 1.13 & 0.27 &$8\times 10^{-9}$\\
ortho-H$_2$CO $2_{11}$--$1_{10}$ & 0.65 & 0.86 & 0.83 & 0.31 & \\
ortho-H$_2$CO $3_{12}$--$2_{11}$ & 0.78 & 1.0 & 0.77 & 1.66 & \\
para-H$_2$CO $1_{01}$--$0_{00}$ & 0.65 & 0.19 & 0.18 & 0.04 &$2\times 10^{-9}$ \\
para-H$_2$CO $2_{02}$--$1_{01}$ & 0.39 & 0.06 & 0.02 & 0.05 & \\
para-H$_2$CO $3_{03}$--$2_{02}$ & 0.82 & 0.90 & 0.54 & 0.99 & \\
para-H$_2$CO $3_{22}$--$2_{21}$ & $<0.12$ & 1.62 & 0.09 & 1.55 & \\
\noalign{\smallskip} 
\tableline
\noalign{\smallskip} 
\multicolumn{6}{c}{Interferometer observations (Jy~beam$^{-1}$, image
peak per channel):}\\
\noalign{\smallskip} 
\tableline
\noalign{\smallskip} 
C$^{18}$O 1--0 & $\lesssim 0.26$ & 1.22 & 0.14 & 1.93 & \\
$^{13}$CO 1--0 & 1.56 & 1.77 & 0.82 & 4.76 & \\
HCO$^+$ 1--0 & 1.30 & 1.11 & 0.92 & 3.77 & \\
\noalign{\smallskip} 
\tableline
\multicolumn{6}{p{14cm}}{$\rm ^a$ intensity entire 5.0 \kms\ component, including ridge;
 taken from Table~\ref{t:decomp}}\\
\multicolumn{6}{p{14cm}}{$\rm ^b$ modeled intensity for disk/envelope system; no depletion assumed}\\
\multicolumn{6}{p{14cm}}{$\rm ^c$ modeled intensity envelope; infinite depletion disk}\\
\multicolumn{6}{p{14cm}}{$\rm ^d$ modeled intensity disk; infinite depletion envelope}\\
\end{tabular}}
\end{table*}

We use the accelerated Monte-Carlo method of \citet{hoge00} to
calculate the molecular excitation and the resulting line emission in
the various beams, including a realistic sampling of spatial scales
recovered by the interferometer observations. Because the disk and
envelope exhibit a range of densities and temperatures, we cannot use
the much simpler escape probability approach here.  We calculate the
line intensities for three different cases (Table~\ref{t:predict}):
for the disk alone, for the envelope alone, and for the combination of
disk+envelope taking into account optical depth effects. The predicted
line intensities are meant to be illustrative only. The general trends
are robust, however.  Adopting `fiducial' molecular abundances derived
for the two foreground clouds (Table~\ref{t:phys}), we find that the
emission of C$^{18}$O observed with OVRO is overestimated by at least
a factor of 5 by the disk+envelope model. This can be accounted for by
adopting a depletion factor of 5-10 for CO in the disk.  A large CO
depletion factor has been seen in many other disks as well
\citep{dutr97, shup01}. From the single-dish lines it can also be
concluded, in particular from CS 7-6, that a significant depletion in
the disk is needed (Table~\ref{t:phys}).  Furthermore, we find that
the higher excitation transitions are dominated by the disk+envelope,
while the lower excitation lines can only be partially explained by
this model.  A fair fraction of these lines must therefore originate
in the ridge.

\subsection{The Dense Ridge}

As mentioned above, Elias~29 appears to be embedded in a ridge traced
by the 1.3~mm continuum and \hcop\ 3-2 line emission, but slightly
offset from the crest of the ridge. How much of the line emission
measured in the single-dish and interferometer beams at 5.0
km~s$^{-1}$ can be attributed to the Elias~29 disk+envelope system,
and how much originates in this dense ridge? This question is
relevant, because it addresses the issue of how much additional column
density resides in the ridge that can harbor the ice-absorption
bands. When we subtract the predicted emission of Elias~29's disk and
envelope from the single-dish line fluxes at 5.0 km~s$^{-1}$ of
Table~\ref{t:decomp}, we can use the ratios of the remaining
intensities to roughly constrain the conditions in the
ridge. Fig.~\ref{f:phys}c shows the results for the C$^{18}$O 1-0/3-2
and \hcop\ 1-0/3-2 lines. The density is constrained at $(4\pm
2)\times 10^5$ cm$^{-3}$, significantly higher than that of the
foreground clouds, while the temperature is found to be similarly low
at $15\pm 5$~K. These parameters are consistent with the appearance of
the ridge in the 1.3~mm and \hcop\ 3-2 maps (cold and dense). The
corresponding column density along the ridge is $1\times 10^{22}$
cm$^{-2}$, comparable to that of the largest column density foreground
cloud at 3.8 km~s$^{-1}$. Elias~29 is offset from the maximum column
density in the ridge, however, and the pencil beam traced in the
infrared absorption features may contain a significantly smaller
column density. The fact that the foreground clouds and, to a lesser
extent, the envelope can already fit the 10 $\mu$m absorption would
suggest that the ridge does not contribute significantly.  This
conclusion does not change if we assume the higher envelope mass of
0.3 $M_\odot$.

\subsection{Location and Thermal History of Ices toward Elias~29}

We have identified four regions of significant column density in the
line of sight of Elias 29: the 2.7 and 3.8 \kms\ foreground clouds,
the dense ridge, and the envelope.  All of these are cold
environments, in which icy mantles can exist, and together they must
account for the rich spectrum of ices seen with the ISO satellite
\citep{boog00b}.  Their relative contribution to the visual extinction
and ice bands depends on which fraction of the column is actually in
front of Elias 29.  The 2.7 and 3.8 \kms\ clouds are likely in front
(\S 4.1) with columns corresponding to visual extinctions $A_{\rm
V}$=2.9 and 8.2.  The visual extinction of the dense ridge, $A_{\rm
V}\leq$5.9, is an upper limit beause the ridge is somewhat displaced
from Elias 29. For the envelope, $A_{\rm V}$ is 2.5 magn. for the
minimum, and 6 magn. for the maximum envelope case. These are upper
limits because the system is seen at low inclination and much of the
envelope material is likely not in front of Elias 29.  The extinction
from either ridge or envelope is however not negligible, given the
absorption seen at 5.0 \kms\ in the CO 6-5 emission line
(Fig.~\ref{f:sdish}). Finally, extinction by the circumstellar disk is
not significant due to its low inclination.

There is observational evidence that ices in molecular clouds form
only above a certain threshold extinction value ($A_{\rm th}$;
\citealt{whit88, tana90}), due to photodesorption at the cloud edges
\citep{tiel82}. Above $A_{\rm th}$, the ice abundances grow linearly
with $A_{\rm V}$. The visual extinction can thus, in principle, be
used as an indicator for H$_2$O ice growth and destructions
mechanisms.  Unfortunately, the usefulness of this relation is limited
by the uncertainty in $A_{\rm th}$.  Measurements of a variety of
sight-lines in $\rho$ Oph indicate $A_{\rm th}$=12 \citep{teix99b},
but $A_{\rm th}<7$ toward OB stars tracing ices in foreground clouds
only \citep{shup00}. The $A_{\rm V}$ values used to calibrate the
aforementioned relation are highly uncertain as well \citep{teix99b}.
While our emission line method has its own uncertainties (beam
dilution, disk/envelope ratio), the generally used H--K color method
may be uncertain by a factor of 2, for example due to an unknown color
of the intrinsic light source.  Indeed, for Elias 29 a much lower
$A_{\rm V}<29$ was found from J--H color (Th.P. Greene, priv. comm.),
compared to the H--K color ($A_{\rm V}=47$; \citealt{lada83}).
Nevertheless, using the relation in \citet{teix99b}, and assuming that
the threshold extinction for H$_2$O formation ($A_{\rm th}$=12)
applies only once for all clouds combined, we find that the total
$A_{\rm V}<23$ corresponds to $N({\rm H_2O})<11\times 10^{17}$ \sqcm,
which is $<$30\% of the observed ice column density \citep{boog00b}.
In this picture, the observed ice column is thus much higher than that
expected from the total visual extinction along the line of sight.
Given the abovementioned uncertainties in the ice column versus
$A_{\rm V}$ and $A_{\rm th}$ relation, the most reliable conclusion
that can be drawn from this is that H$_2$O ice sublimation does not
play an important role in the Elias 29 line of sight.

From a different perspective, the low temperature of the clouds along
the line of sight (Table~\ref{t:phys}) indicates that indeed no
sublimation or other forms of thermal processing
(e.g. crystallization) of polar, H$_2$O--rich ices can take place.  A
similar conclusion was previously drawn from the (low) far infrared
color index, which is found to be a good indicator of thermal
processing of ices \citep{boog00a, tak00}. Even the temperature of the
warmest absorbing component, the inner part of the envelope ($\sim$45
K), is less than the H$_2$O crystallization ($\sim 60$ K) and
sublimation temperatures ($\sim 90$ K). This explains the observed
lack of crystallization signatures in the absorption band profiles of
solid H$_2$O and CO$_2$ \citep{boog00b}.

Thermal processing must have played a role for apolar, CO--rich ices
toward Elias 29, however. The solid CO/H$_2$O ratio is 5\%, which is a
factor of five less than in several other lines of sight within the
$\rho$ Oph cloud \citep{shup00}.  We find that Elias 29 has heated its
envelope sufficiently (17--44 K; \S 4.2) to evaporate CO--rich ices
($T_{\rm subl}\sim$15--20 K; \citealt{tiel91}), but preserve
H$_2$O--rich ices. The envelope, however, harbors only a small
fraction of the ices, and it is unlikely that this 38 $L_{\odot}$
object can heat ices located beyond the envelope, such as the dense
ridge and the two foreground clouds.  Nevertheless, the temperatures
of the gas and dust in the quiescent foreground clouds are similar to,
or slightly higher (Table~\ref{t:phys}) than the sublimation
temperature of apolar ices.  Several known luminous B stars and the
Sco OB2 association play an important role in heating the dust in the
$\rho$ Oph cloud \citep{gree89}.  The differences of the solid
CO/H$_2$O ratio within $\rho$ Oph may thus be explained by the
location of the ices with respect to these luminous heating sources.
Bright external heating sources are absent in the Taurus molecular
cloud, and indeed the CO/H$_2$O ratios are significantly larger in
that cloud \citep{teix99b}.

\subsection{Depletion}

The depletion factor of CO (defined as solid/[gas+solid]) amounts to a
few percent, assuming the ices are present in several of the clouds
toward Elias 29 (using $N$[CO-ice]=$1.7\times 10^{17}$ \sqcm;
\citealt{boog00b}). Direct measurements of solid CO yield larger
depletion factors in the quiescent Taurus molecular cloud (8-40\%;
\citealt{chia95}). Indirect measurements, that rely on gas phase
abundances only, found CO depletion factors as high as 90\% toward
deeply embedded class 0 sources such as NGC 1333: IRAS 4A and 4B
(\citealt{blak95}). We will compare gas phase abundances of other
molecules toward Elias 29 with those of class 0 sources, as well as
the well studied dense cloud TMC~1 (Table~\ref{t:phys}).

Even when taking the uncertainty of gas phase abundances into account,
we find that CO, CS, \formalde, and possibly \hcop\ are an order of
magnitude more depleted in the class 0 sources compared to the clouds
in front of Elias 29. This can be ascribed to the much larger
densities and low dust temperatures in the outer regions ($\gtrsim$
700 AU) of these class 0 objects, and thus shorter depletion times
scales \citep{blak95}.  Depletion is much lower in the TMC~1 cloud
compared to class 0 objects (Table~\ref{t:phys}).  In fact, the
\formalde\ and \methanol\ abundances are an order of magnitude larger
than toward Elias 29. Whether these species are depleted toward Elias
29 cannot be answered directly, because the upper limit to the ice
abundances are typically a few$\times 10^{-6}$ (\citealt{boog00b};
taking $N{\rm (H_2)}=2.9\times 10^{22}$ \sqcm).  This is four and two
orders of magnitude larger than the gas phase \methanol\ and
\formalde\ abundances respectively, and demonstrates that a direct
determination of depletion factors by using infrared ice bands suffers
from a lack of sensitivity.  Alternatively, the difference in gas
phase \methanol\ and \formalde\ abundances between these clouds may
have its origin in different chemical evolutionary states, such was
found within the TMC 1 cloud itself for several carbon bearing species
\citep{prat97}.  Indeed, \formalde\ abundances of several other
quiescent and star forming cores (2--10$\times 10^{-9}$;
\citealt{dick99}) are in much better agreement with Elias 29.

In some class 0 sources, the \methanol\ and \formalde\ abundances and
temperatures are larger than in our comparison source NGC 1333: IRAS
4, possibly because they are more evolved \citep{dish95, blak95}.
This warm gas originates from evaporated ices, that are enriched in
hydrogenated molecules by grain surface reactions during the collapse
of the cold envelope.  Shocks \citep{dish95} and thermal evaporation
\citep{cecc01} in the inner envelope have been proposed as ice
desorption mechanisms, which can enhance the \formalde\ abundance at
radii $<150$ AU by 2 orders of magnitude.  The single dish
observations of Elias 29 do not show evidence of such warm gas, as is
illustrated in Fig.~\ref{f:meth} by comparing the \methanol\ spectrum
with the class 0 source IRAS 16293--2422.  This may well be an effect
of the larger column and accretion rate of the younger class 0
objects.  Infrared absorption line studies do reveal, however, that
significant amounts of warm CO and \water\ are present on small scales
near the protostar, partly related to the outflow, and probably partly
present in the accreting disk's atmosphere (\citealt{boog00b};
A.C.A. Boogert et al., in preparation).  Emission from this gas is
severely diluted in the large single dish milimeter wave beams, and
thus difficult to detect.

Finally, the \hcop\ abundance toward Elias 29 is comparable to that
found toward class 0 objects, but an order of magnitude lower compared
to TMC 1. The importance of depletion is more difficult to establish,
because the \hcop\ abundance is highly dependent on local cosmic ray
flux and electron density. Even in the direction of Elias 29 the
\hcop\ abundance varies dramatically. Although the \hcop\ 1--0 and CS
2-1 lines have similar critical densities, the ridge (at 5.0 \kms) is
much more prominent in \hcop\ than in CS, with respect to the 2.7 and
3.8 \kms\ foreground clouds (Fig.~\ref{f:sdish}). We stress however
that the prominent appearance of the ridge in the \hcop\ 3-2 map
(Fig.~\ref{f:hcop}), is {\it both} an effect of increased \hcop\
abundance and increased density: the ridge is weaker in the column
density tracer \eighteenco, and stronger in the density tracer CS,
relative to the foreground clouds.

\vbox{
\begin{center}
\leavevmode 
\hbox{%
\epsfxsize\hsize
\includegraphics[angle=90, scale=0.47]{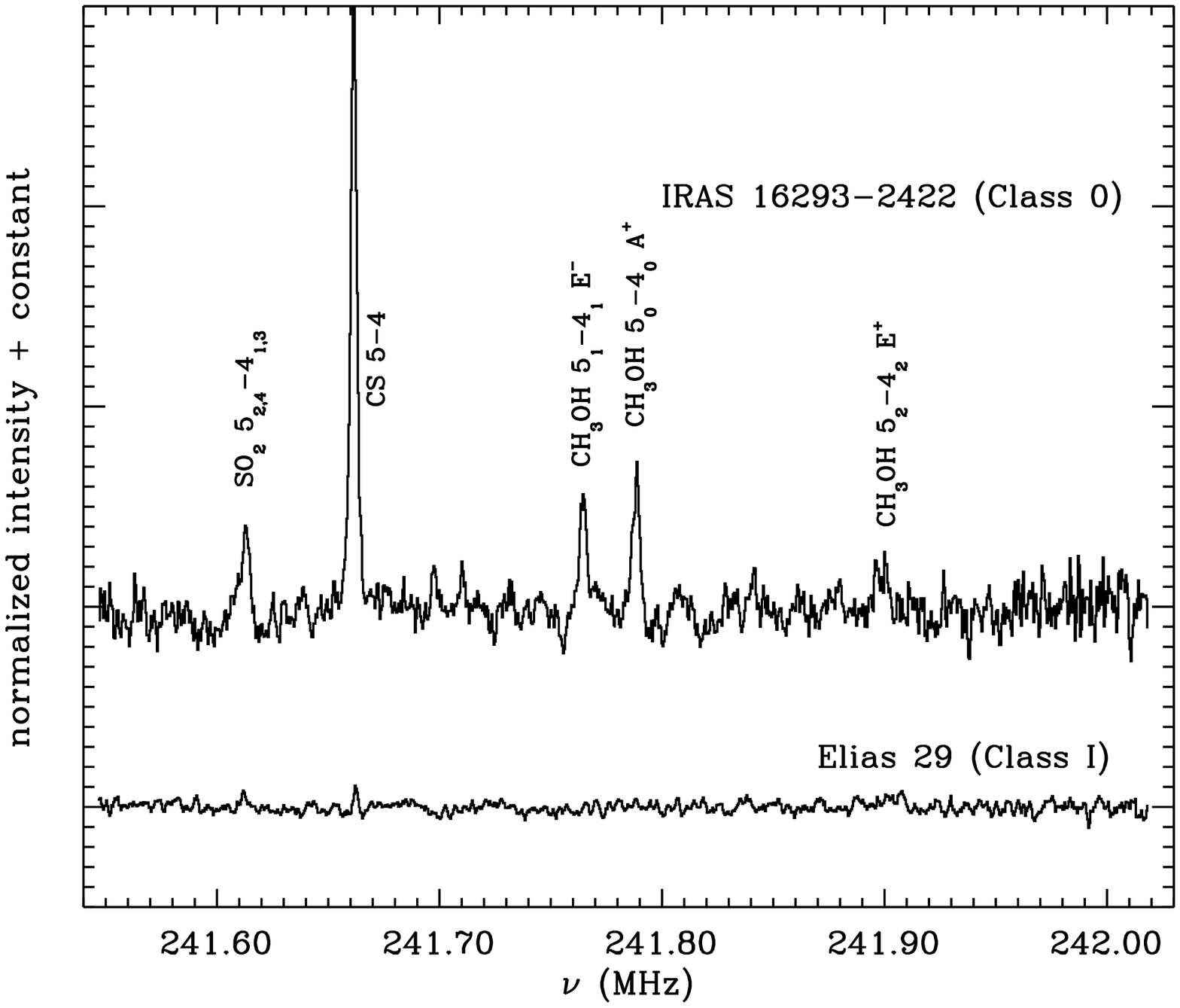}}
\figcaption{Low resolution spectrum showing the enormous difference in
line intensity between the class 0 object IRAS 16293--2422 and the
class I object Elias 29, in all the lines of SO$_2$, CS, and
CH$_3$OH.~\label{f:meth}}
\end{center}}

\section{Summary, Conclusions and Future Work}

We have analyzed infrared and millimeter wave line and continuum
observations to construct a model of the class I protostar Elias 29
and its environment. This model has to contain a number of different
components (summarized in Fig.~\ref{f:schem}): a disk to account for
the $\sim$2--50 \mum\ SED, an envelope contributing to the emission at
1.3 mm (in particular its size), a dense ridge from which Elias 29 may
have condensed, and foreground material which provides most of the
extinction. Elias 29 can then be well described by a 500 AU radius
face-on flared disk with a mass of 0.012 $M_\odot$, embedded in a 6000
AU radius, 0.12 $M_\odot$ envelope.  This large disk provides the
simplest explanation for the observed flat SED, weakly detected 3 mm
continuum emission, and 400 AU radius 5 \mum\ thermal continuum
emission. The present data does however not fully exclude models with
smaller disks. The minimum possible disk has a 30 AU radius and a mass
of 0.002 $M_\odot$ surrounded by an envelope of 0.33 $M_\odot$.  In
this case the combination of disk and envelope emission produces a
flat SED.

The entire system is embedded in a long, dense, cold, and \hcop--rich
ridge. Elias~29 is slightly offset from the crest of this ridge.  In
front of the disk, envelope, and ridge system are two foreground
clouds at a few \kms\ lower radial velocities that cover the entire
field of view. The large column of the foreground clouds,
corresponding to $A_{\rm V}\sim$11, may be responsible for the `class
I' appearance of Elias~29, which would otherwise appear as a T~Tauri
or Herbig Ae/Be star (i.e., optically visible). These same foreground
clouds are also the most likely repository of most of the ices seen
along the line of sight ($\sim 70$\%).  The low temperature of the
foreground clouds explains the observed absence of crystallized ices,
and the presence of large abundances of polar, H$_2$O--rich ices, i.e.
thermal processing did not play a major role for the ices toward Elias
29.  The foreground cloud temperature (25$\pm$15 K) could, however, be
high enough to explain the low abundance of apolar, volatile CO--rich
ices, presumably due to the proximity of a number of luminous B type
stars.  The important question as to whether the ices in the disk or
envelope have experienced thermal processing, as is seen in the
envelopes of massive objects \citep{boog00a, tak00}, cannot be
addressed given the large column of foreground material and the
face-on nature of the system.

This work shows the value of spectrally and spatially resolved
information offered by single-dish and interferometric molecular gas
observations in interpreting infrared ISO satellite observations of
ices along a pencil beam. It shows that, at least for the $\rho$ Oph
cloud, it is crucial to disentangle the different physical components
along the line of sight, since otherwise incorrect conclusions may be
derived, for example, on the origin and evolution of interstellar and
circumstellar ices.

Follow up observations with sensitive submillimeter interferometers
(e.g. ALMA) or high frequency ($>400$ GHz) single dish telescopes with
small beams will clearly provide essential information on the
structure, abundances, and depletion factor of species in the disk and
outflow of Elias 29, and its relationship to the envelope and
ridge. In turn, such studies would provide invaluable information on
the initial conditions of planet formation.

\acknowledgments

We thank Remo Tilanus, Goeran Sandell, and Fred Baas for carrying out
part of the JCMT observations in service mode.  The research of FM and
ACAB at the Caltech Submillimeter Observatory is funded by the NSF
through contract AST-9980846. The research of MRH is supported by the
Miller Institute for Basic Research in Science.

\end{document}